\documentclass{iopjournal}

\usepackage{amsmath,amssymb,amsfonts}
\usepackage{graphicx}
\usepackage{algorithm,algorithmic}
\usepackage{subcaption}
\usepackage{siunitx}

\newcommand{\Ket}[1]{\left|#1\right\rangle}

\begin{document}

\articletype{Article type} %	 e.g. Paper, Letter, Topical Review...

\title{QCxSimulation: Scatter-Aware X-Ray Projection Radiography via Discrete-Time Quantum Walks}

\author{Anja Heim$^{1,*}$\orcid{0000-0002-3670-5403},
        Theobald Fuchs$^{1}$\orcid{0009-0002-5653-3706}}
        Thomas Lang$^1$\orcid{0000-0001-5939-3919},
        Dimitri Prjamkov$^{1}$\orcid{0009-0001-2471-4345},
        Kilian Dremel$^{1}$\orcid{0009-0007-6359-8758}
        Stefan Kasperl$^{1}$\orcid{0000-0002-8118-7609},
        and Christoph Heinzl$^{2,1}$\orcid{0000-0002-3173-8871}

\affil{$^1$Fraunhofer Institute of Integrated Circuits IIS, 90768 F\"{u}rth, Germany}

\affil{$^2$University of Passau, 94032 Passau, Germany}

\affil{$^*$Author to whom any correspondence should be addressed.}

\email{anja.heim@iis.fraunhofer.de}


\begin{abstract}
X-ray projection radiography is a key non-invasive imaging technique used in medical diagnostics and industrial inspection. The simulation of X-ray projections is commonly used to optimise acquisition protocols and improve image quality before performing costly and radiation-intensive scans. Classical photon transport simulations that include realistic X-ray scattering physics are computationally expensive because they require the sampling of a large number of distinct scattering paths. This limits the practical exploration of parameter spaces such as beam energy, source geometry or object composition. Quantum computing offers the potential to solve high-dimensional problems faster by making use of quantum properties such as superposition.
This work introduces a discrete-time quantum walk algorithm that simulates the transport of X-ray photons through heterogeneous volumes. It approximates the physics of X-ray projection radiography, including processes such as photoelectric absorption and higher-order scattering, including Compton and Rayleigh scattering. The quantum walk encodes all admissible photon paths into a single quantum state, enabling all scattering histories to be propagated simultaneously via the superposition principle. This quantum state representation enables flexible readout of various imaging modalities, including the primary, i.e., unscattered, image, or images exclusively containing Rayleigh and Compton scattering of specified orders.
A quantitative comparison with classically computed reference simulations shows that the proposed quantum walk accurately reproduces radiographic projections, given the limitations of the underlying physical model. These results indicate that quantum circuits for X-ray transport can produce accurate radiographic images and imply that, as quantum hardware scales up, these algorithms could outperform classical Monte Carlo-based approaches in large-scale, scatter-aware virtual imaging studies.

\end{abstract}

\section{Introduction}
X-ray imaging is critical for clinical~\cite{luo_artificial_2026, velleman_state---art_2026} and industrial applications~\cite{hassan_hybrid_2026, metiner_deep_2025}, but it is susceptible to scattering effects that can degrade image quality~\cite{anagnostou_effect_2026, andriiashen_quantifying_2023}. Optimising scan parameters such as acceleration voltage, tube current and specimen orientation is necessary to mitigate these artefacts. Physical experimentation to determine optimal settings is infeasible due to cost and safety constraints. X-ray simulation overcomes this limitation by computing virtual projections from phantom data, such as surface meshes and material properties. This enables the cost-effective exploration and optimisation of diverse scan settings, while also enabling the quantification of scattering effects.

Simulating higher-order X-ray scattering requires the computationally demanding task of tracking all photon trajectories and scattering paths. We propose using quantum computing to address this challenge by introducing the following: (1) a quantum state encoding scheme representing individual X-ray photon trajectories and scattering events (Rayleigh and Compton), and (2) a quantum algorithm exploiting quantum superposition to track all potential scattering paths simultaneously, enabling efficient higher-order scattering computation.

% TODO: Add how the paper is structured

\subsection{X-ray Radiography Simulation}
\begin{figure*}
  \centering
  \begin{subfigure}{0.32\textwidth}
    \centering
    \includegraphics[width=\linewidth]{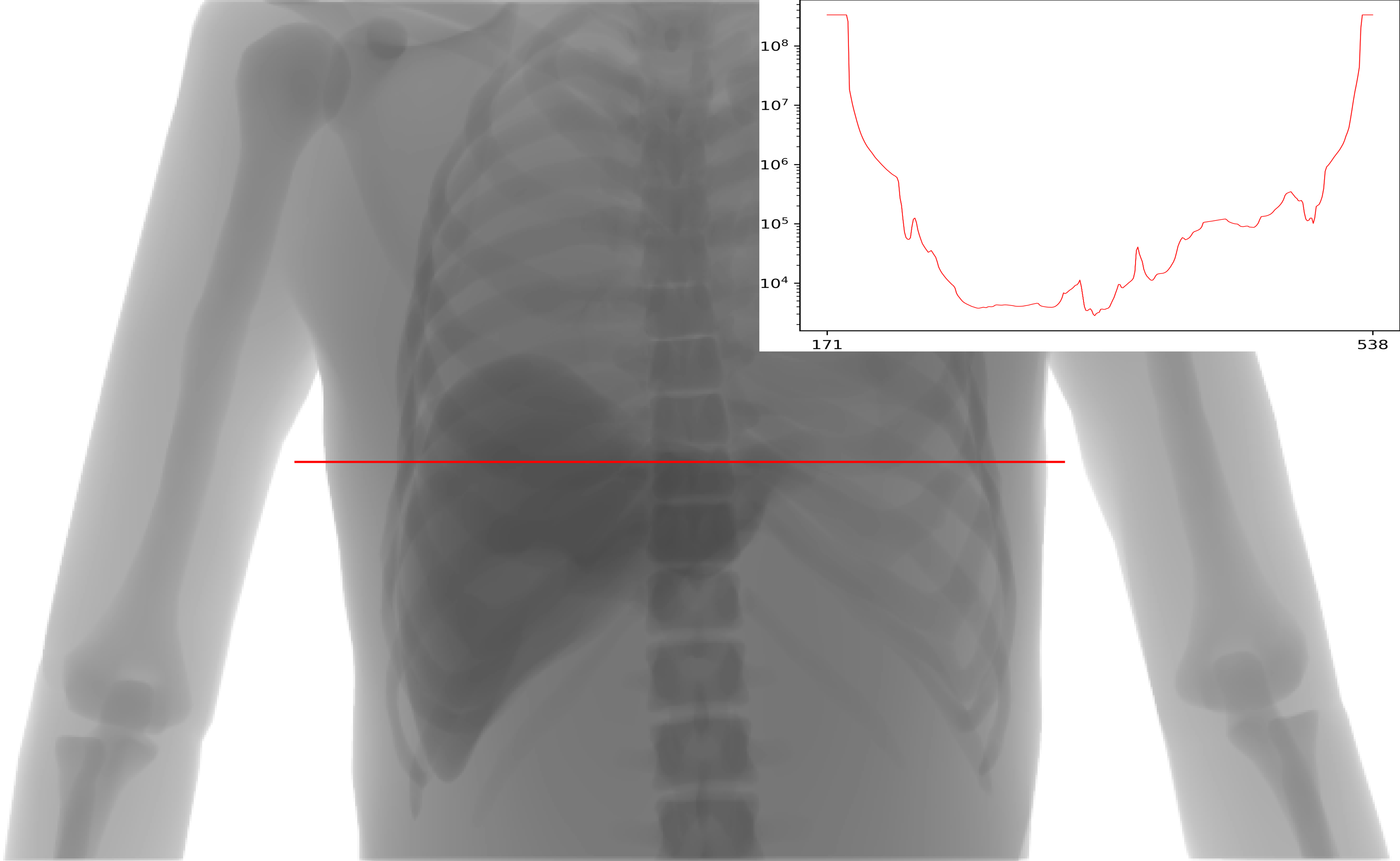}
    \caption{Primary radiation image}
    \label{fig:scattering_ideal}
  \end{subfigure}
  \hfill
  \begin{subfigure}{0.32\textwidth}
    \centering
    \includegraphics[width=\linewidth]{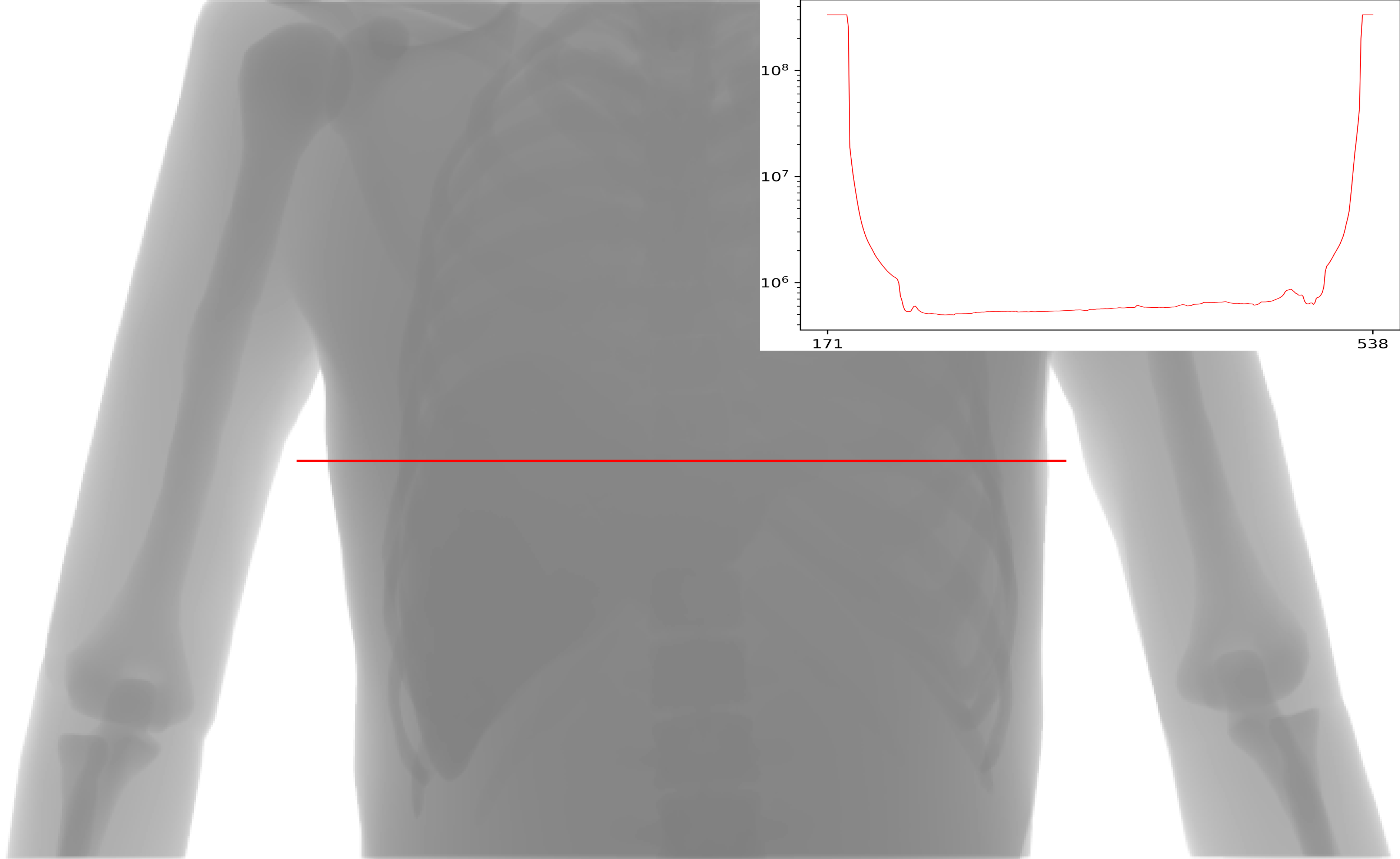}
    \caption{Added first-order scattering}
    \label{fig:scattering_1st}
  \end{subfigure}
  \hfill
  \begin{subfigure}{0.32\textwidth}
    \centering
    \includegraphics[width=\linewidth]{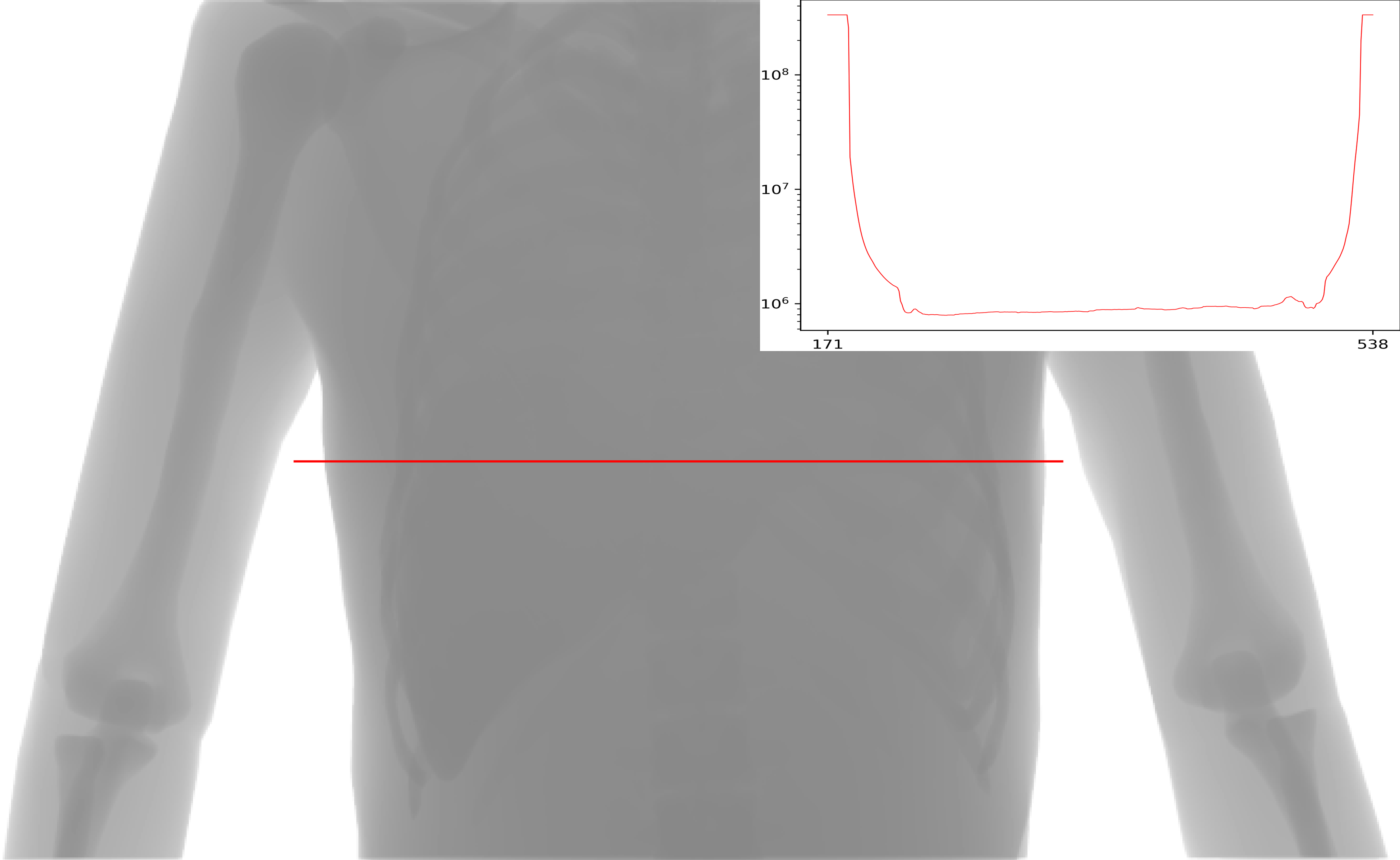}
    \caption{Added second-order scattering}
    \label{fig:scattering_2nd}
  \end{subfigure}
  \caption{X-ray radiography simulations of a torso phantom. (\protect\subref{fig:scattering_ideal}) shows the primary radiation image generated by simulating X-ray photon absorption, while (\protect\subref{fig:scattering_1st}) and (\protect\subref{fig:scattering_2nd}) include X-ray scattering effects up to first and second order, respectively. X-ray scattering severely impacts tissue contrast and many fine structural details are lost.}
  \label{fig:scattering_motivation}
\end{figure*}

X-ray projection radiography is widely used in medical imaging and industrial non-destructive testing, where virtual X-ray projections are crucial for optimizing acquisition protocols and image quality. \autoref{fig:scattering_motivation} illustrates simulated projections of a region of interest from the \emph{mesh50\_XCAT} phantom~\cite{auer_mesh_2023}. The projections were generated with the Fraunhofer IIS \emph{XSimulation} tool~\cite{makarov_computed_2023} in a typical clinical setting with a tube voltage of 80 kV and a tube current of 200 mA. 

When only primary radiation is modelled, i.e. photoelectric absorption without scatter, the resulting projection (c.f \autoref{fig:scattering_ideal}) exhibits high contrast and clearly delineates different tissue classes, as confirmed by the intensity profile along the indicated line. However, X-ray scattering contributes substantially to the detected signal at these energies. Photons may undergo elastic (Rayleigh) scattering~\cite{strutt_xv_1871} or inelastic (Compton) scattering~\cite{compton_quantum_1923}, which redistributes the intensity across the pixels of the detector. Including single scattering events (\autoref{fig:scattering_1st}) reduces the contrast in regions of large thickness, such as the thorax, where many primary photons are deflected from their original path. Adding second-order scattering (\autoref{fig:scattering_2nd}) further reduces contrast, particularly around the fifth and sixth rib pair. It also erodes low-contrast soft-tissue detail in the grey-value profile. Therefore, accurate modelling of these higher-order scattering processes is critical for realistic radiographic simulation. However, such modelling significantly increases the computational cost. For this example, \emph{XSimulation} requires only \SI{57}{\milli\second} for the primary-only image, but about \SI{28}{\second} for first-order and \SI{23}{\minute} for second-order scatter for a single projection, illustrating the steep cost of higher-order scattering.

Existing radiography simulators can be grouped into Monte Carlo, analytical, and hybrid approaches. Monte Carlo transport, as implemented in GEANT4~\cite{agostinelli_geant4simulation_2003} and GPU-accelerated variants~\cite{badal_accelerating_2009}, yields the most faithful physics by tracking individual photon histories, but is computationally expensive. Analytical simulators employ fast GPU ray tracing to calculate primary transmission using line integrals~\cite{vidal_simulation_2009}. They typically handle scattering with approximations such as spatially varying blurring~\cite{gallio_gpu_2015} or explicit first-order models based on differential cross sections and additional rays~\cite{reiter_simct_2016}. Extending these schemes to higher orders becomes prohibitive because the number of potential scattering paths grows exponentially. To our knowledge, \emph{XSimulation}~\cite{makarov_computed_2023} is the only analytical tool that models up to second-order scatter. \emph{XSimulation} was validated against GEANT4 simulations in terms of accuracy, and reportedly is much faster than GEANT4 simulations for the same settings~\cite{schielein_analytische_2018}. Hybrid methods combine analytical primary images with Monte Carlo-estimated scatter~\cite{oconnell_hybrid_2025,tabary_sindbad_2007}. This process often involves simulating fewer photons and applying smoothing to obtain realistic scatter fields~\cite{neffati_novi-sim_2023,bellon_artist_2007}. All three families exhibit a rapid increase in runtime with specimen size and scatter order. This severely limits large parameter space explorations and virtual CT studies. 
The primary challenge in these computations is the tracking of all potential photon paths, i.e. rays, whose number increases exponentially with the number of potential scatterings per photon. Quantum computing promises significant increases in processing speed due to the unique superposition principle~\cite{von_neumann_mathematische_1932}, which enables the simultaneous tracking of all photon paths. 

\subsection{Quantum Computing}
Quantum computing is particularly well suited to modelling wave-like transport processes such as X-ray photon propagation, due to its ability to directly represent complex probability amplitudes in a finite-dimensional Hilbert space. An $n$-qubit register spans a $2^n$-dimensional state space with orthonormal computational basis states $\{\Ket{k}\}_{k=0}^{2^n - 1}$. A general pure state has the form 
\begin{equation}
\Ket{\psi} = \sum_{k=0}^{2^n - 1} \alpha_k \Ket{k}, \quad \sum_k |\alpha_k|^2 = 1
\end{equation}
so that the complex amplitudes $\alpha_k$ encode the full probability distribution over all classical configurations $k$ of the register. The time evolution of a closed quantum system is modelled by unitary operators $U$, i.e. propagators, such that $\Ket{\psi (t + \Delta t)} = U\Ket{\psi(t)}$. This is equivalent to Schrödinger's time evolution in quantum mechanics~\cite{von_neumann_mathematische_1932}. A projective measurement in the computational basis results in a classical outcome $k$ with probability $|\alpha_k|^2$, thereby collapsing the superposition as dictated by Born's rule~\cite{born_quantenmechanik_1926}. For problems that naturally decompose into many alternative, discrete outcomes, such as random walks on a lattice, or combinatorial configurations, a single quantum state can encode the amplitudes of an exponentially large number of these outcomes. Each measurement samples one specific realisation from this range of possibilities.

For problems involving motion or propagation on a discrete structure, it is natural to use quantum walks, the quantum analogue of classical random walks on a graph or grid~\cite{Aharonov2003}. In the discrete-time setting, the Hilbert space is divided into a position register and an register, which represents the different propagation modes and often referred to as the 'coin' space. A basis state can be written as $\Ket{\mathbf{x}_i, \omega_i}$, where $\mathbf{x}_i$ denotes a grid point, e.g. any point on a lattice, and $\omega_i$ encodes a direction, spin, or other internal degree of freedom that determines how the walker moves. A single step of a walk consists of two unitaries, specifically a coin operator and a shift operator.

The coin operator $C$ mixes the internal states at fixed position, 
\begin{equation}
    C\Ket{\mathbf{x}_i, \omega_i} = \sum_{\omega'} c_{\omega'}(\mathbf{x}_i) \Ket{\mathbf{x}_i, \omega'}
\end{equation}
where the coefficients $c_{\omega'}(\mathbf{x}_i)$ encode the local transition amplitudes from the incoming internal state to the outgoing ones at position $\mathbf{x}_i$. This sum is a superposition in itself, considering all coin states $\omega'$. In many applications, these amplitudes are selected to represent an update rule.

The shift operator $S$ translates the position register conditionally on the internal state,
\begin{equation}
S\sum_{\omega'} c_{\omega'}(\mathbf{x}_i) \Ket{\mathbf{x}_i, \omega'} = \sum_{\omega'} c_{\omega'}(\mathbf{x}_i) \Ket{\mathbf{x}_i + \Delta x(\omega'), \omega'}
\end{equation}
where $\Delta x(\omega')$ is the discrete step on the grid associated with the internal label $\omega'$, e.g. a step in the desired direction.

Applying the combined step $SC$ repeatedly constructs a superposition over all paths that are consistent with the discretized geometry and the local transition rules encoded in $C$. From a physical perspective, this is a discrete path-integral representation. Rather than stochastically following one random-walk trajectory at a time, as in classical Monte Carlo simulation of Markov chains, a quantum walk propagates amplitudes for all admissible trajectories simultaneously. Only upon measurement do these superposed paths collapse and one possible path or endpoint is observed, with the probability given by the Born rule.

% Related Work Quantum Ray Tracing
% Zhang2025
% Lu2022
% Lu2023
% Alves2019
Discrete-time quantum walks have been applied to simulate radiative transport through quantum versions of classical ray-based light transport methods. Among the earliest proposals was the work of Lanzagorta and Uhlmann~\cite{lanzagorta_hybrid_2005}, who introduced a hybrid classical-quantum framework in which Grover’s search was employed to perform ray-primitive intersection tests. However, this approach processes each ray independently, offering no improvement in scaling with respect to the number of rays.
Subsequent work has developed three main paradigms for quantum radiative transport. 
% Hybrid Classical–Quantum Ray Tracing with Grover Search 
In hybrid classical-quantum workflows \cite{Alves2019, Santos2025, Zhang2025}, Grover-based search is only used for the most expensive geometric subtask, the visibility test. This involves identifying the first object that a ray hits in the scene. Instead of using a classical computer to check the ray against all $M$ objects one by one, which would require $\mathcal{O}(M)$ intersection tests, the quantum part applies Grover's amplitude-amplification search to find a match in $\mathcal{O}(\sqrt{M})$ oracle calls. Meanwhile, all the other parts of the rendering algorithm, such as tracing the rays' paths through reflections and refractions, run entirely on a classical processor. In practice, the effective speed increase is significantly restricted by the time taken to encode the scene geometry into a reversible 'oracle' circuit, as well as by the noise and circuit depth limitations of current NISQ hardware.
% Fully Quantum Path Tracing with Amplitude Estimation
Alternative approaches utilise fully quantum path tracing to encode many light paths simultaneously in quantum superposition and employ quantum amplitude estimation to read out pixel intensities \cite{Lu2022, Lu2023}. These methods have been shown to achieve superior convergence, with the estimation error scaling as $\mathcal{O}(1/N)$ compared to the classical Monte Carlo rate of 
$\mathcal{O}(1/\sqrt{N})$, where $N$ denotes the number of samples or oracle queries. The fundamental trade-off is that these approaches require significantly deeper quantum circuits and more precise quantum control making them more challenging to implement on currently available quantum devices.
%grid-based quantum transport schemes
Grid-based quantum transport schemes employ quantum walks or lattice Boltzmann methods to efficiently encode radiative transport on discretised grids. Mosier et al.~\cite{Mosier2023} developed quantum ray marching, which uses conditional and coin unitaries to represent an exponentially large superposition of branched light paths with only a polynomial number of qubits, achieving linear computational scaling equivalent to classical methods. In astrophysics, Devkota and Wise~\cite{Devkota2025} implemented a quantum lattice Boltzmann method that stores direction-resolved radiation intensity as quantum amplitudes at grid points. This method has been primarily employed for the purpose of studying weakly scattering radiative transfer in large-scale astrophysical domains. Our X-ray radiography algorithm follows this paradigm, adopting a discrete-time quantum walk on a 3D voxel grid.

\section{Method}
Our work adopts a quantum walk for modelling X-ray photon transport through attenuating and scattering media. We discretise the transport domain into a finite 3D grid and represent each photon by a basis state over position, discrete travel direction, and further describing information. Material-dependent scatter interaction probabilities, such as photoelectric absorption, Compton scattering, and Rayleigh scattering, are encoded in the coin operation constructed from physical theory. The shift operation implements movement and boundary condition checks on the grid. Repeated application of these operations yields a quantum state, whose amplitudes correspond to all possible multi-scatter photon paths consistent with the chosen discretisation. To obtain detector statistics and the resulting radiographs, we perform repeated projective measurements of the quantum walk.
The following sections will highlight the key steps in realizing this strategy within a quantum circuit.

\subsection{Physical Problem and Discretization}

% General Environment
Our algorithm models the transport of monoenergetic X-ray photons through a heterogeneous, voxelized medium $\Omega$, defined as a finite uniform Cartesian grid $\Omega = \{0,\dots,N_{x}-1\} \times \{0,\dots,N_{y}-1\} \times \{0,\dots,N_{z}-1\} \subset \mathbb{Z}^{3}$ with $ \lvert \Omega \rvert = N_xN_yN_z$ voxels in total. Each voxel $\mathbf{x} = (x,y,z) \in \Omega$ corresponds to the physical cube having a uniform voxel edge length of $d=1.0~\mathrm{cm}$. We express the centres of all voxels by coordinate triples $(x,y,z)$, or their equivalent linear indices $i\in\{0,\ldots,N_xN_yN_z\}$ in standard Z-order traversal. This linearised index is encoded in a state $\Ket{\mathrm{pos}}$ in its binary representation. This encoding is more qubit-efficient than a direct tensor product encoding of $x, y, z$, and therefore reduces the overall qubit count and gate depth while still uniquely representing every voxel. 

Photons are injected with a primary energy of $E_0$ in the range of $40-450~\mathrm{keV}$, which encompasses the typical energies used in clinical and industrial X-ray applications. As a photon propagates through a position 
$\mathbf{x}_i\in \Omega$, it is subject to three mutually competing interaction processes, namely photoelectric absorption (P), Compton scattering (C) and Rayleigh scattering (R). The probability of each process occurring depends on the local material and photon energy and is described via
\begin{align*}
    P_j(E,Z) = 1-e^{-\mu_j(E,Z)d}, \quad j\in\{P,C,R\},
\end{align*}
where $d$ denotes the distance of the photon to the emitting source within that material.

% physical effects - 1. photoelectric absorption
In case of \textbf{photoelectric absorption}, the photon is absorbed, i.e. removed from the simulation, and its entire energy is deposited locally in the voxel as dose. This process may result in the emission of secondary electrons, which, for the purposes of this study, will be disregarded. The coefficient $\mu_P(E,Z)$ denotes the the linear attenuation coefficient of the local material ($Z$) at the given energy ($E$) and can be obtained from experimental data.
In case of \textbf{Rayleigh scattering}, the photon scatters of a bound electron elastically, i.e. without transferring energy. Only its direction is affected by some angle depending on the material and the energy. The coefficient $\mu_R(E,Z)$ expresses this via its differential cross section for this scattering type as obtained from experimental data. Since in our relevant energy regime Rayleigh scattering is known to have little influence on the final image, we approximate the scattering angles by a uniform sampling over the full $4\pi$ solid angle.
Differently, \textbf{Compton scattering} is an inelastic scattering process guided by its differential cross section (described by the Klein-Nishina formula~\cite{Klein1929}) yielding respective coefficients $\mu_C(E,Z)$. Due to the inelasticity of this process, some energy of the photon is transferred to the electron, which is ejected from its orbital subsequently. Depending on the solid angle $\theta$ the photon is scattered to, the energy of the photon after scattering is estimated by the Compton energy-angle relation $E'=E/(1+\alpha (1-\cos\theta))$ where $\alpha = E/(m_e c^2)$ is the dimensionless energy parameter with $m_e$ being the electron rest mass. Both the number of Compton scattering events encountered during the simulation and the respective energy loss are encoded in the quantum state in discretized form as described below.

% Material
\paragraph{Material data.} The medium is composed of $N_{\mathrm{mat}}$ distinct materials, with each material $m=1,\ldots,N_{\mathrm{mat}}$ is assumed to be elemental, characterized by its atomic number $Z_m$ and mass density $\rho_m$. For a given energy $E$ and each material, the linear attenuation coefficients $\mu_{P}(E,Z)$, $\mu_{C}(E,Z)$ and $\mu_{R}(E,Z)$ are computed from xraylib~\cite{Brunetti2004}. Xraylib provides mass attenuation coefficients $\widetilde{\mu_j}:=\mu_j / \rho$ as functions of energy and atomic number. These are converted to linear coefficients via $\mu_j(E,Z)= \widetilde{\mu_j}(E,Z) \rho, j \in \{P, C, R\}$. Additionally, direction- and energy-dependent Compton look-up tables are precomputed using xraylib's differential cross sections to model angular and energetic redistribution within the discrete state space.
%
% Spatial Discretization (7 qubits)
    % Linear index: idx=x+5y+25z for x,y,z∈{0,1,2,3,4} (125 voxels)
    % Voxel edge length: δt=1.0 cm (physically scaled)
    % Voxel-to-material assignment: each voxel belongs to at most one material (enforced by validation)
    % Why linearization --> otherwise more qubits required, bad for simulation
%The continuous domain $\Omega$ is discretised using a uniform Cartesian grid consisting of $5 \times 5 \times 5$ cubic voxels, each with an edge length of $d=1.0~\mathrm{cm}$. 

% Explanation M gate
Material heterogeneity is represented by assigning each material $m \in \{0,\ldots,N_{mat} - 1\}$ a mutually disjoint set of voxel indices $V_m \subset \{0,\ldots,124\}$. In the circuit this is realised by a material sampling gate $M_m$ which flips a dedicated ancilla qubit to $\Ket{1}$ if and only if $\mathrm{pos} \in V_m$. Voxel exclusivity ($V_m \cap V_{m'} = \emptyset$ for $m \neq m'$) is enforced at input validation, so that at most one material ancilla is active for any position. This guarantees that a photon interacts with at most one material at any step and avoids double counting of attenuation.
%
% Explanation Shift gate
Transport between voxels is described by a shift operation which increments the linear index by a direction-dependent integer offset $\Delta \mathrm{pos}(d)$ (see direction discretization below). If a proposed shift would move the photon out of the grid (i.e. to some $(x,y,z)\not\in\Omega$), the photon is "killed" and its survival flag in the quantum state is set to "absorbed". A photon that reaches $z=0$ (the detector) is registered as detected. All other photon positions at the end of the simulation represent loss to the environment. For a source located at $z=4$, a photon travelling straight along $-z$ must traverse four voxels to reach the detector. Consequently we use at least five quantum-walk steps to allow for one or more scattering events and still permit arrival at the detector.

\paragraph{Direction Discretization.}
% Direction Discretization (4 qubits, 14 usable directions)
    % This geometry provides 9 forward-going directions (Klein-Nishina favors small angles), 1 backscatter, and 4 lateral directions—totaling the observable angular range
    % Reasons why exactly these 9 forward-going directions
The photon propagation direction is discretised into 14 representative unit vectors on the unit sphere. These are encoded in a 4-qubit direction register $\Ket{\mathrm{dir}}$, which can represent 16 basis states. We use 14 of these states to represent physically meaningful directions, and leave two states unused. Let $d \in \{0,…,13\}$ denote the direction index. The corresponding integer grid displacements $\Delta r_d=(\Delta x_d,\Delta y_d,\Delta z_d)$ per step are given in table \autoref{tab:directions}. These can be grouped into forward-going ($d \in \{4,6,7,8,9,10,11,12,13\}$), backward-going ($d=5$), and laterally moving ($d \in \{0,1,2,3\}$).
\begin{table}[tbh]
\centering
\begin{tabular}{r r r r}
\hline
$d$ & $\Delta r_d$ & Interpretation \\
\hline
0  & $(1,0,0)$     & lateral \\
1  & $(-1,0,0)$    & lateral \\
2  & $(0,1,0)$     & lateral \\
3  & $(0,-1,0)$    & lateral \\
4  & $(0,0,-1)$    & primary beam direction \\
5  & $(0,0,+1)$    & backscatter \\
6  & $(1,0,-1)$    & $\approx 45^\circ$ forward \\
7  & $(-1,0,-1)$   & $\approx 45^\circ$ forward \\
8  & $(0,1,-1)$    & $\approx 45^\circ$ forward \\
9  & $(0,-1,-1)$   & $\approx 45^\circ$ forward \\
10 & $(1,0,-2)$    & $\approx 26.6^\circ$ forward \\
11 & $(-1,0,-2)$   & $\approx 26.6^\circ$ forward \\
12 & $(0,1,-2)$    & $\approx 26.6^\circ$ forward \\
13 & $(0,-1,-2)$   & $\approx 26.6^\circ$ forward \\
\hline
\end{tabular}
\caption{Directions, unit vectors, and interpretations for indices $d=0,\dots,13$.}
\label{tab:directions}
\end{table}
Thus nine of the 14 directions correspond to forward-going motion ($-z$ component), four are purely lateral ($z=0$ component), and one represents backward scattering ($+z$ component). This allocation reflects the strongly forward peaked nature of the Klein-Nishina differential cross section at diagnostic energies~\cite{schielein_analytische_2018}. Most Compton scatters are confined to small or moderate polar angles around the incident beam direction, with comparatively few large-angle backscatter events.
The integer displacements $\Delta r_d$ define the voxel-to-voxel shift during each transport step. Directions with $\Delta z=-1$ move the photon one voxel closer to the detector per step. Those directions with $\Delta z=-2$ move the photon by two voxels per step and therefore model more grazing, small-angle forward paths. Lateral moves, where $\Delta z=0$, capture $90^\circ$ scatters that do not change the longitudinal position.
The continuous scattering angle $\theta$ between an incoming direction $d_{\mathrm{in}}$ and an outgoing direction $d_\mathrm{out}$ is calculated by the dot product
\begin{equation*}
    \theta(d_{\mathrm{in}},d_\mathrm{out}):= \cos^{-1}\left( \langle \Delta\mathbf{r}_{\mathrm{in}},\Delta\mathbf{r}_\mathrm{out}\rangle / (\|\Delta\mathbf{r}_\mathrm{in}\|_2\|\Delta\mathbf{r}_\mathrm{out}\|_2) \right).
\end{equation*}
From these 14 directions we precompute a $14 \times 14$ matrix of scattering angles. To further simplify the calculations, we consider two groups of directions, "forward-like" $B_0(d_\mathrm{in})$ and "deflected" $B_1(d_\mathrm{in})$, explicitly given by
\begin{align*}
  B_0(d_\mathrm{in}) &= \{d_\mathrm{out} \mid \theta(d_\mathrm{in},d_\mathrm{out}) \leq 3\pi/8\},\\
  B_1(d_\mathrm{in}) &= \{d_\mathrm{out} \mid d_\mathrm{out} \not\in B_0(d_\mathrm{in}) \}.
\end{align*}
Considering the "forward-like" scattering angles, the photon path is only slightly altered and hence it loses little energy. The specific angle of $3\pi/8$ is chosen heuristically. In the subsequent calculations, we ignore this energy loss.
%and, for each $d_{in}$, define two angular neighbourhoods, namely, $B_0(d_{in}) = {d_{out}: cos \theta \geq cos 67.5°}$ and $B_1(d_{in}) = {d_{out}: cos\theta < cos 67.5°}$. These two bins correspond to “forward-like” ($\theta \lesssim 67.5°$) and “deflected” ($\theta \gtrsim 67.5°$) scattering, respectively. 
For each material and energy level, xraylib’s differential Compton cross section $d\sigma /d\Omega(E,\theta,Z)$ is integrated over these discrete sets to obtain direction- and energy-dependent probabilities $P(B_b | d_{in}, E_k, Z)$, where $b \in \{0,1\}$. These probabilities are used to implement a unitary redistribution of probability amplitudes from $\Ket{d_\mathrm{in}}$ to favour forward directions.%a coherent superposition over the allowed outgoing directions in $B_0$ or $B_1$.

This discretisation is a compromise between angular resolution and qubit and gate resources. A finer tessellation of the unit sphere would require more than four direction qubits, increasing the circuit width and depth beyond what is currently feasible to compute. The chosen 14 directions sample all qualitatively distinct scattering regimes (small-angle forward, moderate-angle forward, lateral and backward) with higher resolution in the forward cone dictated by the shape of the cross section. For imaging observables such as total transmission and scatter-to-primary ratios, this level of angular discretisation is sufficient to reproduce the leading-order anisotropy of Compton scatter while keeping the quantum circuit resource requirements moderate.

\paragraph{Energy discretization.}
% Energy Discretization  (2 qubits, 4 levels)
     % Energy level k∈{0,1,2,3} represents the photon's degradation state after deflected Compton scatters
     % Physical justification: The 67.5° threshold separates forward-like scatters (ΔE≈0, no level decrement) from deflected scatters (ΔE≈1 level). This approximation captures the dominant Klein-Nishina energy transfer;
The photon energy is represented by a two qubit register $\Ket{energy}$, yielding four discrete energy levels, labelled $k=0,1,2,3$. Level $k=3$ corresponds to the source energy $E_0$. Lower levels represent progressively energy-degraded photons after one or more large-angle Compton scatters. The mapping from the abstract levels $k$ to physical energies $E_k(d)$ is constructed in a precomputation step, separately for each direction $d$, using xraylib’s Compton energy and differential cross section. For a given material and source energy $E_0$, we define $E_3(d) = E_0$. Then, for each lower level $k=0,1,2$ and each incoming direction $d_{\mathrm{in}}$, we compute an expected value for the energy after one additional “deflected” Compton scatter (i.e. with $\theta \in B_1(d_{\mathrm{in}})$) by
\begin{equation}
    E_k(d_{\text{in}}) =
\frac{
\displaystyle \sum_{d_{\text{out}} \in B_1(d_{\text{in}})}
w\bigl(E_{k+1}, \theta_{d_{\text{in}} \to d_{\text{out}}}, Z\bigr)\,
E'\bigl(E_{k+1}, \theta_{d_{\text{in}} \to d_{\text{out}}}\bigr)
}{
\displaystyle \sum_{d_{\text{out}} \in B_1(d_{\text{in}})}
w\bigl(E_{k+1}, \theta_{d_{\text{in}} \to d_{\text{out}}}, Z\bigr)
}.
\end{equation}
where $E'(E,\theta)$ is the Compton formula above, and $w(E,\theta,Z)$ is the Klein-Nishina differential cross section including the incoherent scattering function provided by xraylib. This yields a direction-dependent cascade of typical energies $E_3(d)\rightarrow E_2(d) \rightarrow E_1(d) \rightarrow E_0(d)$. For a representative case of aluminium at $E_0=80~\mathrm{keV}$ the forward direction $d=4$ yields typical level energies of the order $E_3 \approx 80~\mathrm{keV}$, $E_2 \approx 65~\mathrm{keV}$, $E_1 \approx 50~\mathrm{keV}$, $E_0 \approx 40~\mathrm{keV}$. The precise values used in the simulation are obtained by integrating the Klein-Nishina differential cross section with xraylib and may differ slightly from these rounded figures, but the quoted numbers correctly indicate the scale of the energy degradation captured by the four-level discretization at $80~\mathrm{keV}$.

During the quantum walk, the energy level is updated in a coarse-grained manner. Each Compton event into the deflected angular bin 
$B_1$, i.e. $\theta(d_\mathrm{in},d_\mathrm{out}) \geq 67.5^\circ$, decrements the $\Ket{\mathrm{energy}}$ register by one level (unless already at $k=0$), while forward-like scatters leave it unchanged. A separate binary counter tracks the total number of Compton scatters of any angle for diagnostic purposes, so that the energy level captures the number of large-angle, energy-degrading events and the Compton counter captures the total scatter order. To assign physically consistent interaction probabilities to each level, the attenuation coefficients $\mu_j^{(k)}(Z)=\mu_j(E_k^{(ref)},Z)$, where $j \in {P,C,R}$, are precomputed once via xraylib at the representative energy $E_k^{(ref)}$ of each level $k$ and stored in a look-up table. During the walk, the appropriate entry is selected according to the photon's current level $k$. This construction is essential to capture the qualitatively different energy dependence of photoelectric absorption, Compton scattering and Rayleigh scattering within the small number of discrete energy states.

The choice of four levels is motivated by the limited number of significant energy-degrading events in the geometries considered. With six transport steps and centimetre-scale voxels, typical photon histories contain at most 3 deflected Compton scatters before the photon is either absorbed or exits the domain. Four levels are therefore sufficient to resolve the main spectral evolution from $E_0$ to a substantially degraded energy. Increasing the number of energy levels would require additional qubits and substantially more multi-controlled rotations, which would enlarge the circuit beyond the current capabilities of simulators, with limited impact on the considered observables of interest (total transmission, scatter fractions, energy-resolved projection).

In summary, the combination of a modest but physically motivated 3D spatial grid, a direction set that resolves forward, lateral and backward scattering with finer sampling in the forward cone, and  a four-level energy cascade derived from xraylib, yields a discretisation that reproduces the dominant features of X-ray transport and Compton-dominated scattering in a resource-efficient way that is compatible with current quantum simulators.

\subsection{X-ray Simulation: Per-Step Quantum Circuit}
The X-ray transport described in Section 2.1 is implemented as a discrete-time quantum walk. Each step applies the unitary operator 
\begin{equation}
    U_{\text{step}} = S\left( \prod_{m=0}^{N_{\text{mat}}-1} M_m^{-1} C_m M_m \right),
\end{equation}
consisting of material-local interaction blocks followed by a global shift.

\textbf{The quantum state registers} encoding a photon are $\Ket{\mathrm{pos}}$, $\Ket{\mathrm{dir}}$, $\Ket{\mathrm{energy}}$ and $\Ket{\mathrm{alive}}$, holding the 7-qubit linear voxel index, the 4-qubit direction index, the 2-qubit energy level, and a 1-qubit survival flag, respectively. A small number of ancilla qubits implement angular binning, interaction flags and boundary detection within each step but carry no persistent physical state.

\textbf{The material operation $M_m$} compares for each material $m$ the $\Ket{\mathrm{pos}}$ register with the pre-defined voxel set for that material. It flips a dedicated ancilla to $\Ket{1}$ wherever the photon resides in that material. Since the voxel sets are mutually exclusive, at most one material ancilla is active in any basis state. Applying $M_{m}^{-1}$ after the interaction block restores all ancillas, preventing residual entanglement.

\textbf{The coin operation $C_m$} is conditioned on the material ancilla set by $M_m$ and the current energy level $k$. It applies the three interaction processes, photoelectric absoprtion, Rayleigh scattering and Compton scattering, whose transition amplitudes are drawn from precomputed xraylib look-up tables. Photoelectric absorption is implemented as a controlled rotation that transfers amplitude from the $\Ket{\mathrm{alive} = 1}$ to the $\Ket{\mathrm{alive} = 0}$ branch with probability 
$P_P(\mu_{P}^{(k)},Z_m)$. For Compton scattering, the outgoing direction is sampled in two stages: an ancilla qubit is first rotated to select the angular class 
$B_0$ or $B_1$ according to the precomputed Klein-Nishina weights, after which the direction register is transformed, conditioned on this class, into a uniform superposition over all outgoing directions in $B_0(d_{\mathrm{in}})$ or $B_1(d_{\mathrm{in}})$. A deflected scatter with $\theta \geq 67.5^\circ$ additionally decrements $\Ket{\mathrm{energy}}$ by one level, unless $k=0$. Rayleigh scattering redistributes the direction register uniformly over all 14 directions while leaving the energy register unchanged.
To record the scattering history without additional persistent qubits, a pair of per-step ancilla qubits are flagged whenever a Compton or Rayleigh event occurs, then immediately measured and written to a classical bitstring. This yields a classical scatter log per step and per material while leaving the coherent registers unaffected.

\textbf{The shift operation $S$} increments $\Ket{\mathrm{pos}}$ by the direction-dependent offset $\Delta\mathrm{pos}(d)$, after all material blocks have been applied. Before shifting, it checks whether the resulting position would lie outside $\Omega$. If so, it sets $\Ket{\mathrm{alive} = 0}$ and leaves $\Ket{\mathrm{pos}}$ unchanged. Photons reaching the detector face $z=0$ are flagged as detected, all other out-of-bounds exits are recorded as lost.

% I would add a schematic qubit state representation at the beginning of this subsection
% so, something like:   |pos, alive, Compton, ... >

% Sumary
Within a single time step, the circuit therefore operates as follows: for each material in the scene, the algorithm (i) tags the portion of the quantum state located in that material via $M_m$, (ii) applies the appropriate physics-driven interaction $C_m$ to that tagged subspace, and (iii) removes the tag via $M_{m}^{-1}$, restoring the material flags so they can be safely reused. After all materials have been processed in this way, the global shift $S$ advances all surviving photon amplitudes through the geometry and applies the boundary conditions. Repeating this composite step operator realises a discrete-time, coined quantum walk on the coupled position-direction-energy state space that encodes X-ray transport, with the walk’s local "coin" unitaries fully determined by the underlying interaction cross sections.

% Measurement and observables
After the prescribed number of steps, all quantum registers are measured once. Each measurement shot yields a single bitstring containing the final position, the final direction and discrete energy level, as well as the survival status and the Compton and Rayleigh flags for each material at each step.
Aggregating many such shots (e.g. $10^4 - 10^5$) produces the usual CT observables: primary and scatter projections on the detector, scatter-to-primary ratios, and energy-resolved images. From the CT perspective, the quantum walk therefore plays the role of a stochastic transport kernel whose transition probabilities are fixed by xraylib cross sections and the chosen discretisation. The quantum circuit details ensure that this kernel is implemented unitarily and efficiently on a quantum simulator approximating the underlying X-ray physics. The following section will show different example applications and qualitatively show the physical accuracy of our proposed method.

\section{Results} \label{sec:results}
% Look for the use cases in Paper_ResultsSection.pptx starting from slide 16.
% General QC Simulation
All simulation results presented in this section are obtained on a $5 \times 5 \times 5$ voxel grid with a voxel edge length of $d=1.0~\mathrm{cm}$. %The three unused basis states ($\mathrm{pos}=125,126,127$) are never populated by the shift dynamics and thus do not affect the physical results. 
The plane $z=4$ is the source plane, the plane $z=0$ is the detector plane, and the outer faces at $x,y,z \in \{0,4\}$ are the boundaries \cite{schielein_analytische_2018}. 
% Explanation for Detector size
The $5 \times 5$ detector provides sufficient spatial resolution to resolve a central object region together with a possible surrounding air margin, which is adequate for all configurations studied. Each simulation uses six quantum walk steps, which is the minimum number required to allow a photon emitted at $z=4$ to reach the detector at $z=0$ via the primary beam direction while still permitting several intermediate scattering event.
% Simulation runs
All runs are performed with $10,000$ measurement shots, a value that balances statistical precision against computational cost for the observables of interest. The shots can be interpreted as the photons which are sent out from the source, one after the other.
% Use case intro
Three use cases are considered. The first is a study of an aluminium phantom, in which three qualitatively different source geometries are compared. The second use case involves a two-material phantom consisting of adjacent aluminium and copper blocks. The third use case is a superposition study, in which the detector projection obtained from a $3 \times 3$ source is compared with the normalised sum of nine independent simulation results from a single source.

% Explanation of result images +
% Definition of primary, compton 1x, compton 2x, all comopton and all detected
The simulation results are presented as detector projection images, where each pixel records the number of photons that reached the detector plane. To characterise different physical interaction histories, the detected photons are decomposed into scatter-order categories: \textit{Primary} photons have traversed the phantom without any scattering event, travelling exclusively in the forward direction ($d=4$). \textit{Compton~1x} and \textit{Compton~2x} contain photons that experienced exactly one or two Compton scattering events, respectively and \textit{all~Compton} accumulates all photons with at least one Compton interaction. The \textit{all~detected} image is the inclusive sum over all photons that reached the detector, regardless of their scattering history. For each source configuration, these five projections are displayed side by side showing the corresponding photon counts as a colour-coded heatmap.

\subsection{Use Case 1} \label{sec:usecase1}
\begin{figure}[tbh]
 \centering
    \includegraphics[width=\textwidth]{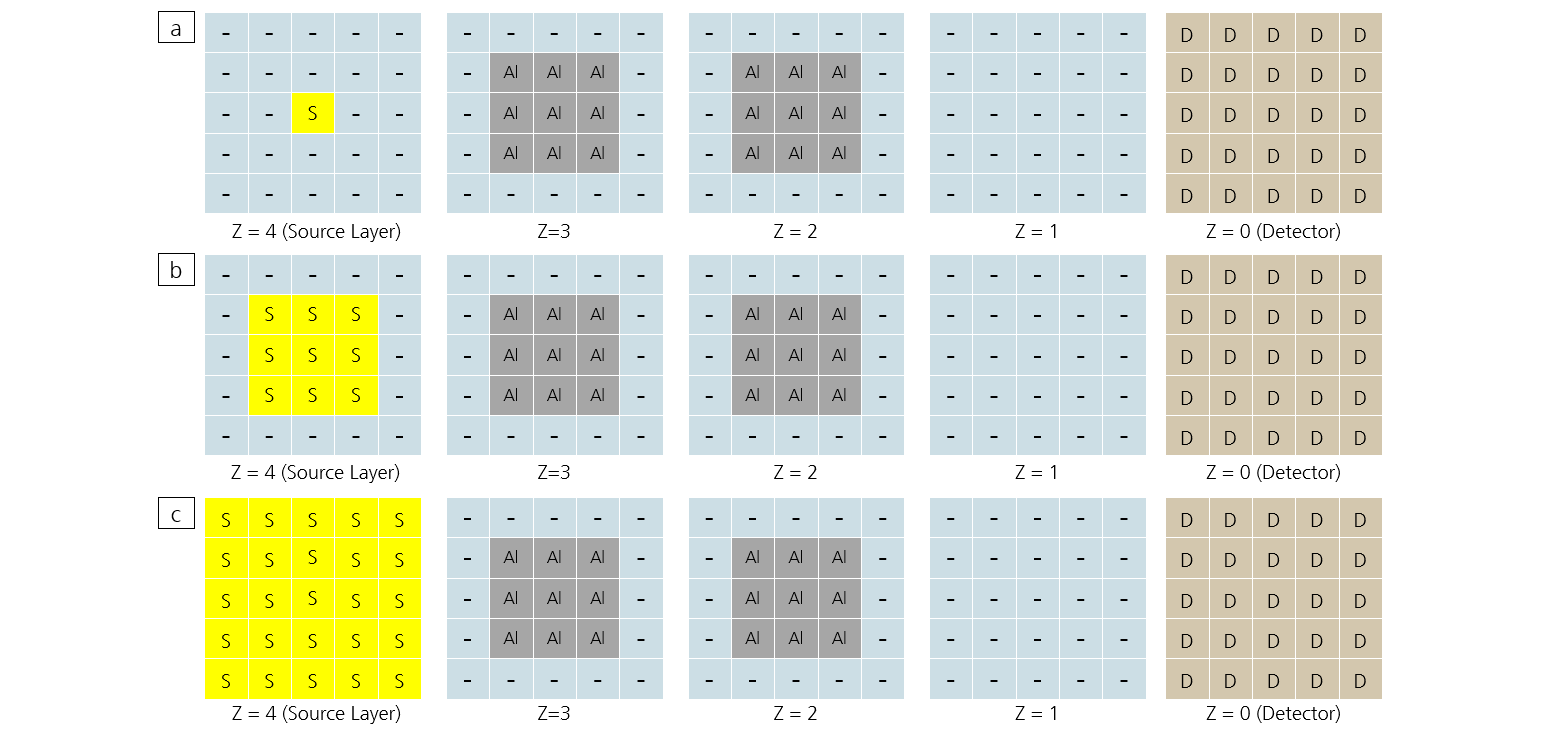}
 \caption{Use Case 1 Structure: Three different light source geometries with a $3 \times 3 \times 2$ aluminium block. (a) represents a point source, (b) a $3 \times 3$, and (c) a full-field source.}
\label{fig:usecase1_intro}
\end{figure}

% see the 3 simulations in QCxSimulation\test_algorithms
% simulation using different sources
The algorithm is first validated using an aluminium phantom ($Z=13$, $\rho=2.7~\mathrm{g/cm^3}$) consisting of a $3 \times 3 \times 2$ voxel block centred in the grid. Three different source geometries are simulated, which are a single point source located at grid position $(2, 2, 4)$, a $3 \times 3$ source covering the central nine positions at $z = 4$, and a full-field source uniformly illuminating the entire $5 \times 5$ source plane. \autoref{fig:usecase1_intro} shows the simulation structure. All three simulations use a photon energy of $80~\mathrm{keV}$. 
%
% what the sources physically describe
The three source configurations serve different purposes. The point source focuses the beam on a single voxel, producing a primary beam whose scattered light can be accurately quantified. The $3 \times 3$ source uniformly illuminates the entire material block, exercising all nine voxel positions simultaneously and probing the spatial uniformity of the coin operator across the phantom. The full-field source illuminates all 25 positions at the source layer, including regions outside the material block. This also enables testing of the algorithm's behaviour in air voxels, where virtually no interaction occurs.
%
% Why is this use case physically interesting
% Computation:
% Pfadlänge L=1 cm (1 Voxel, z.B. Rand des Blocks):
    % T = e^{−0.136*1} = e^{−0.136} = 0.8729 = 87%
% Pfadlänge L=2cm (2 Voxel, Strahl durch z \in {2,3}):
    % T = e^{−0.136*2} = e^{−0.272} = 0.7620 = 76%
With an energy of $80~\mathrm{keV}$ in aluminium, the simulation is performed in a range, where neither full absorption nor full transmission dominates. The photoelectric and Compton contributions are both significant at this energy, and the $3 \times 3 \times 2$ geometry presents an effective path length of up to $2~\mathrm{cm}$ through the material, yielding a transmitted fraction on the order of $76\%-87\%$ depending on the photon path. This phantom therefore constitutes a meaningful test for the interplay between absorption and scattering within the quantum walk.

\begin{figure}[tbh]
 \centering
\includegraphics[width=\textwidth]{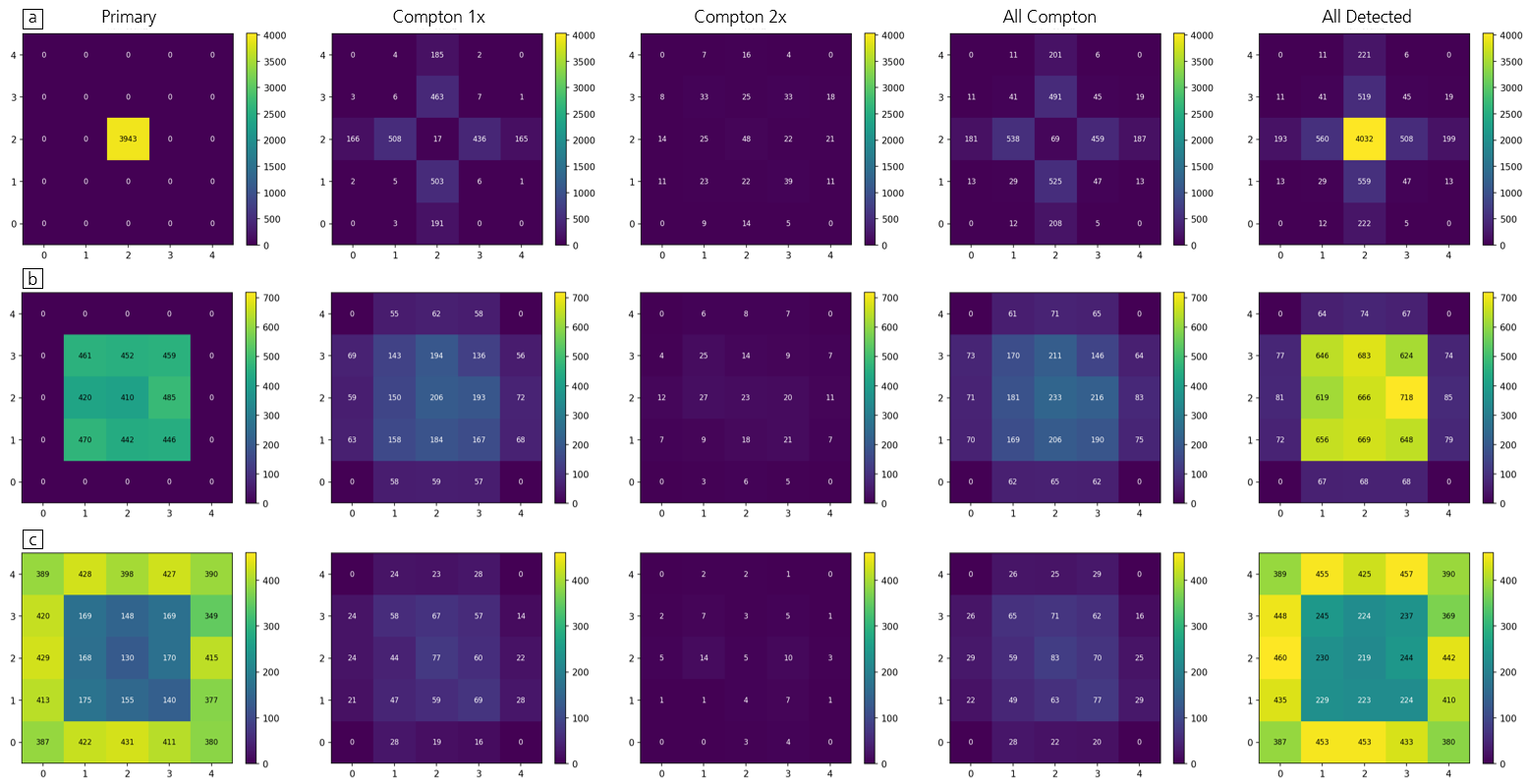}
 \caption{Photon count of detector projections for Use Case 1: each subfigure corresponds to a scatter-order category showing photon counts. (a) point source, (b) $3 \times 3$ source, (c) full-field source}
\label{fig:usecase1_result1}
\end{figure}

\begin{figure}[htbp]
    \centering
    % First row: two subfigures side by side
    \begin{subfigure}[b]{0.49\textwidth}
        \centering
        \includegraphics[width=\textwidth]{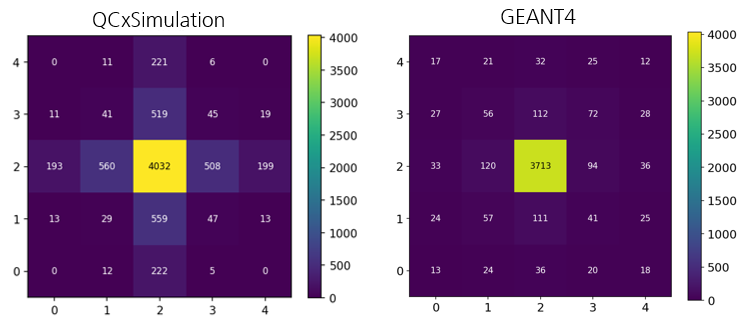}
        \caption{Scenario \autoref{fig:usecase1_intro}~(a)}
        \label{fig:sub1}
    \end{subfigure}
    \hfill
    \begin{subfigure}[b]{0.49\textwidth}
        \centering
        \includegraphics[width=\textwidth]{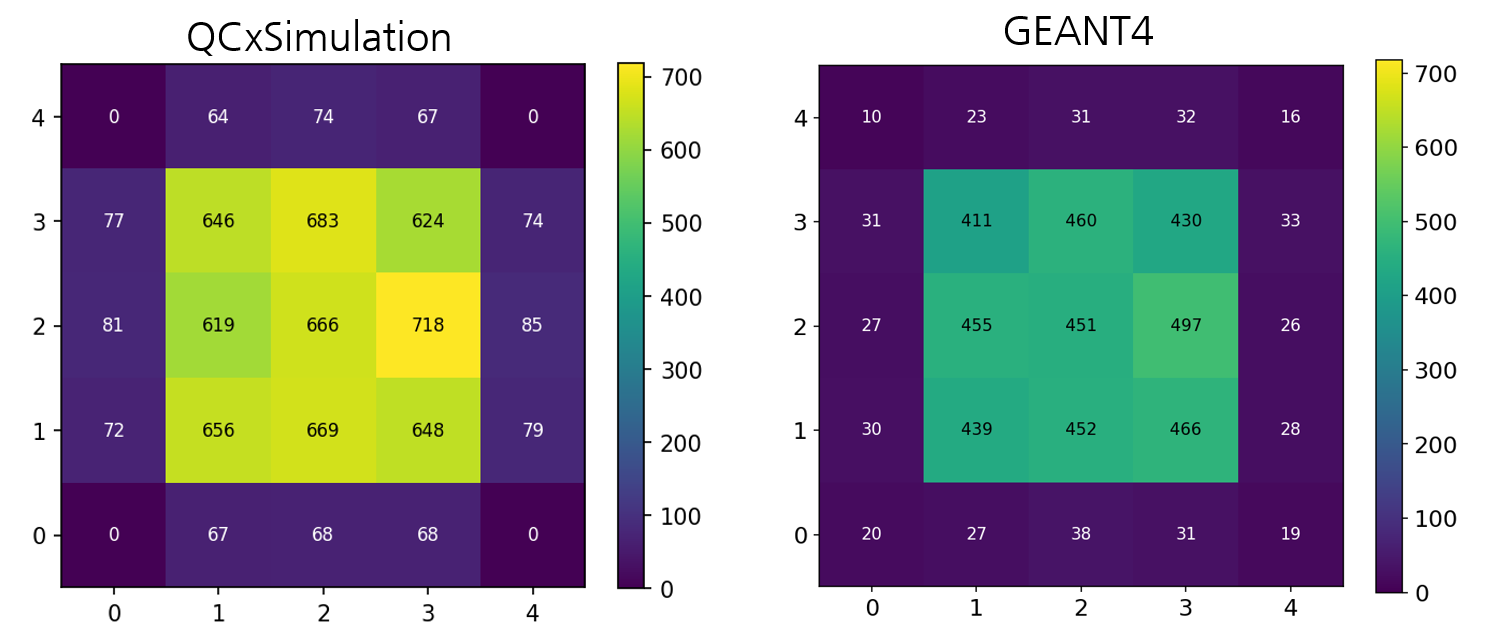}
        \caption{Scenario \autoref{fig:usecase1_intro}~(b)}
        \label{fig:sub2}
    \end{subfigure}

    % Second row: one subfigure centered
    \vspace{0.15cm}
    \begin{subfigure}[b]{0.49\textwidth}
        \centering
        \includegraphics[width=\textwidth]{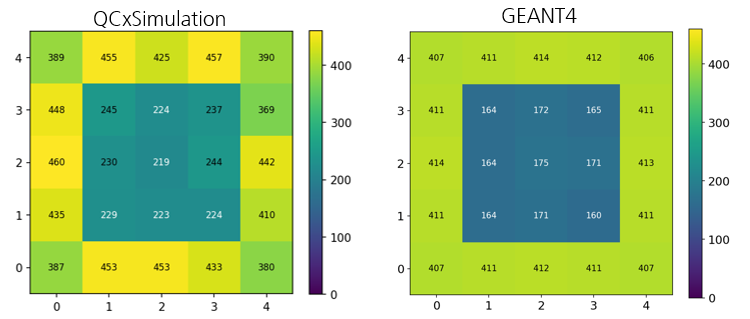}
        \caption{Scenario \autoref{fig:usecase1_intro}~(c)}
        \label{fig:sub3}
    \end{subfigure}

    \caption{Photon count of detector projections for Use Case 1: Comparison of QCxSimulation results \textit{all-detected} with the results from GEANT4}
    \label{fig:usecase1_geant4}
\end{figure}

% describing the result images
\autoref{fig:usecase1_result1} shows the resulting detector projections for all three source configurations. For the point source in \autoref{fig:usecase1_result1}(a), the primary signal is confined to the central pixel $(2,2)$, with Compton contributions spreading into neighbouring pixels. The $3\times 3$ source in \autoref{fig:usecase1_result1}(b) produces a spatially uniform primary signal over the central $3 \times 3$ detector region. The full-field source in \autoref{fig:usecase1_result1}(c) additionally illuminates the surrounding air voxels. 
%
% Geant4 Comparison
To validate the quantum walk results against an established reference, the three source configurations are additionally simulated in the GEANT4 framework~\cite{agostinelli_geant4simulation_2003}. The GEANT4 model uses an identical parallel-beam geometry, in which photons are injected at the centre of the source plane voxels analogously to the quantum model. The detector is configured as a CdTe photon-counting detector that registers only the first photon hit per primary particle. The GEANT4 simulation employs the same total number of photons, i.e. 10,000 photons. After the simulation, the continuous detector plane is discretised into $1~\mathrm{cm} \times 1~\mathrm{cm}$ cells matching the quantum walk voxel grid, and the hit counts are accumulated in a $5 \times 5$ two-dimensional histogram. This yields a spatially resolved reference projection that is directly comparable to the \textit{all-detected} images produced by the quantum circuit.
\autoref{fig:usecase1_geant4} shows the \textit{all-detected} detector projection of the quantum walk simulation alongside the corresponding GEANT4 reference for each of the three source configurations. The spatial intensity distributions demonstrate a high degree of qualitative similarity across all three cases: the primary shadow of the aluminium block, the lateral spread of the scattered signal, and the uniform response in air regions are consistently reproduced. Deviations in both the total number of photons and the spatial distribution of the scattering contributions are to be expected, given that the physical system had to be significantly simplified in order to be encoded on a quantum circuit. In particular, the coarse direction set of 14 discrete vectors, the four-level energy discretisation, and the voxel-scale step length approximate the continuous angular and spectral distributions of the full GEANT4 physics. These approximations result in slight differences in the intensities of the scatter patterns, but do not change the main spatial structure of the projections. This confirms that the quantum walk captures the dominant radiographic behaviour of the aluminium phantom.

\subsection{Use Case 2}
\begin{figure}[tbh]
 \centering
    \includegraphics[width=\textwidth]{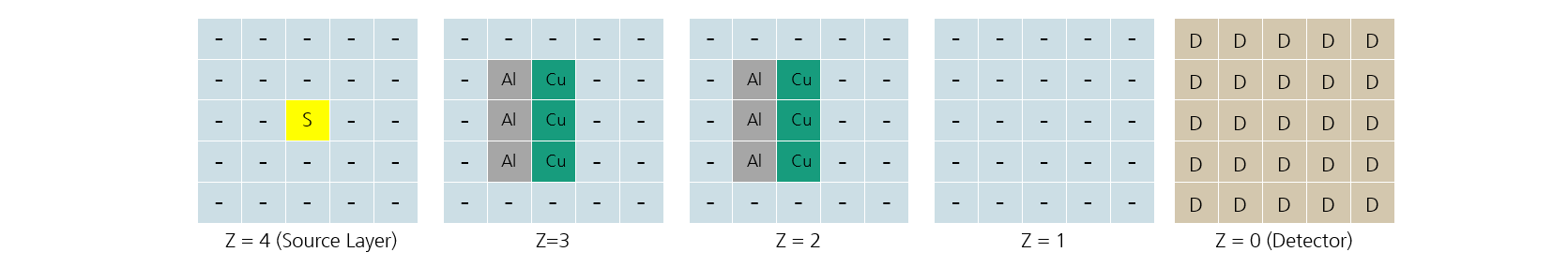}
 \caption{Use Case 2 Structure: Two-material phantom
consisting of an aluminium and a copper block, both centred in the grid at $y \in \{1,2,3\}$, $z\in\{2,3\}$. A point source is placed at $(2,2,4)$.}
\label{fig:usecase2_intro}
\end{figure}

% see the 3 simulations in QCxSimulation\test_asymmetric_Alu_Cu
% multi-material at different keVs
The second use case introduces a two-material phantom to test the algorithm's ability to distinguish materials with markedly different atomic numbers, see \autoref{fig:usecase2_intro}. The phantom consists of two adjacent $1 \times 3 \times 2$ voxel slabs centred in the grid: an aluminium block ($Z=13$, $\rho=2.7~\mathrm{g/cm^3}$) occupying $x=1$, $y \in \{1,2,3\}$, $z \in \{2,3\}$, and a copper block ($Z=29$, $\rho=8.96~\mathrm{g/cm^3}$) occupying $x=2$, $y\in \{1,2,3\}$, $z \in \{2,3\}$. A single point source is placed at grid position $(2,2,4)$, i.e. directly above the copper block, so that the primary beam traverses the copper block while the aluminium block occupies the immediately adjacent column. Three photon energies are simulated, which are $80~\mathrm{keV}$, $220~\mathrm{keV}$, and $400~\mathrm{keV}$.

\begin{figure}[tbh]
 \centering
    \includegraphics[width=\textwidth]{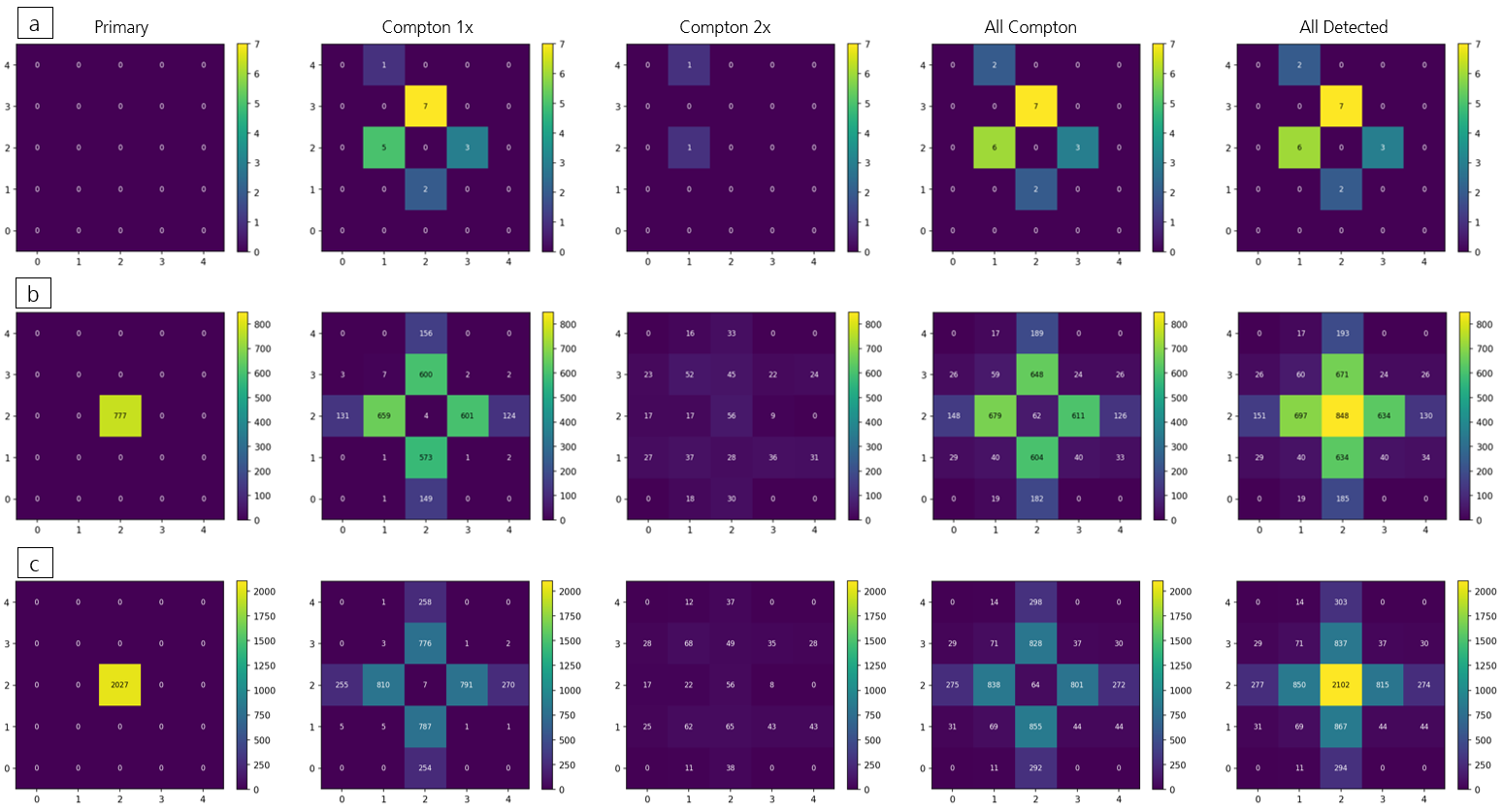}
 \caption{Photon count of detector projections for Use Case 2: (a)~$80~\mathrm{keV}$: the primary signal at pixel $(2,2)$ is strongly suppressed by photoelectric attenuation in copper, while Compton contributions spread laterally into the adjacent aluminium column at $x=1$. (b)~$220~\mathrm{keV}$: Compton scattering becomes dominant, increasing primary transmission through copper and broadening the scatter distribution. (c)~$400~\mathrm{keV}$: the cross section difference between aluminium and copper is significantly reduced, yielding a more uniform scatter distribution.}
\label{fig:usecase2_result1}
\end{figure}

% Why is this use case physically interesting
The three energy levels probe qualitatively distinct interaction ranges. At $80~\mathrm{keV}$, photoelectric absorption dominates for copper, while aluminium attenuates only weakly a strong material contrast between the two columns. At $220~\mathrm{keV}$, the photoelectric contribution diminishes substantially for both materials and Compton scattering becomes dominant, reducing but not eliminating the contrast. At $400~\mathrm{keV}$, Compton scattering dominates for both materials and the cross-section difference between aluminium and copper is significantly reduced, yielding a low-contrast, scatter-dominated regime. This energy progression therefore evaluates the sensitivity of the quantum walk to material-dependent attenuation and scatter signatures across a wide spectral range. Furthermore, the lateral asymmetry of the phantom, with the primary beam at $(2,2,4)$ incident on the copper slab and the aluminium slab displaced by one voxel in the x direction, tests whether the coin operator produces the expected asymmetric detector shadow.
%
% describing the result images
\autoref{fig:usecase2_result1} shows the resulting detector projections for all three energy configurations. The spatial asymmetry of the phantom is clearly reflected across all three results. Because the point source is located at $(2,2,4)$, directly above the copper slab
$x=2$, the primary beam traverses the copper column exclusively, while the aluminium slab at $x=1$ is not directly illuminated and contributes only through Compton-scattered photons that are deflected laterally from the copper column. This produces a characteristic asymmetric scatter signal, in which the \textit{Compton~1x} projection extends into the $x=1$ column whilst the $x=3$ air region remains dark.
The energy dependence of this pattern is pronounced. At $80~\mathrm{keV}$ in \autoref{fig:usecase2_result1}~(a), photoelectric absorption dominates in copper, strongly suppressing the primary signal at pixel $(2,2)$, and only a small number of scattered photons reach the detector. At $220~\mathrm{keV}$, in \autoref{fig:usecase2_result1}~(b), the photoelectric absorption diminishes and Compton scattering becomes more dominant, resulting in a visibly brighter primary pixel at $(2,2)$ and a broader scatter distribution into the aluminium column. At $400~\mathrm{keV}$ in \autoref{fig:usecase2_result1}~(c), the Compton-dominated range yields the brightest primary signal and a more uniform scatter distribution, reflecting the reduced cross section difference between aluminium and copper at higher energies.

\begin{figure}[htbp]
    \centering
    % First row: two subfigures side by side
    \begin{subfigure}[b]{0.49\textwidth}
        \centering
        \includegraphics[width=\textwidth]{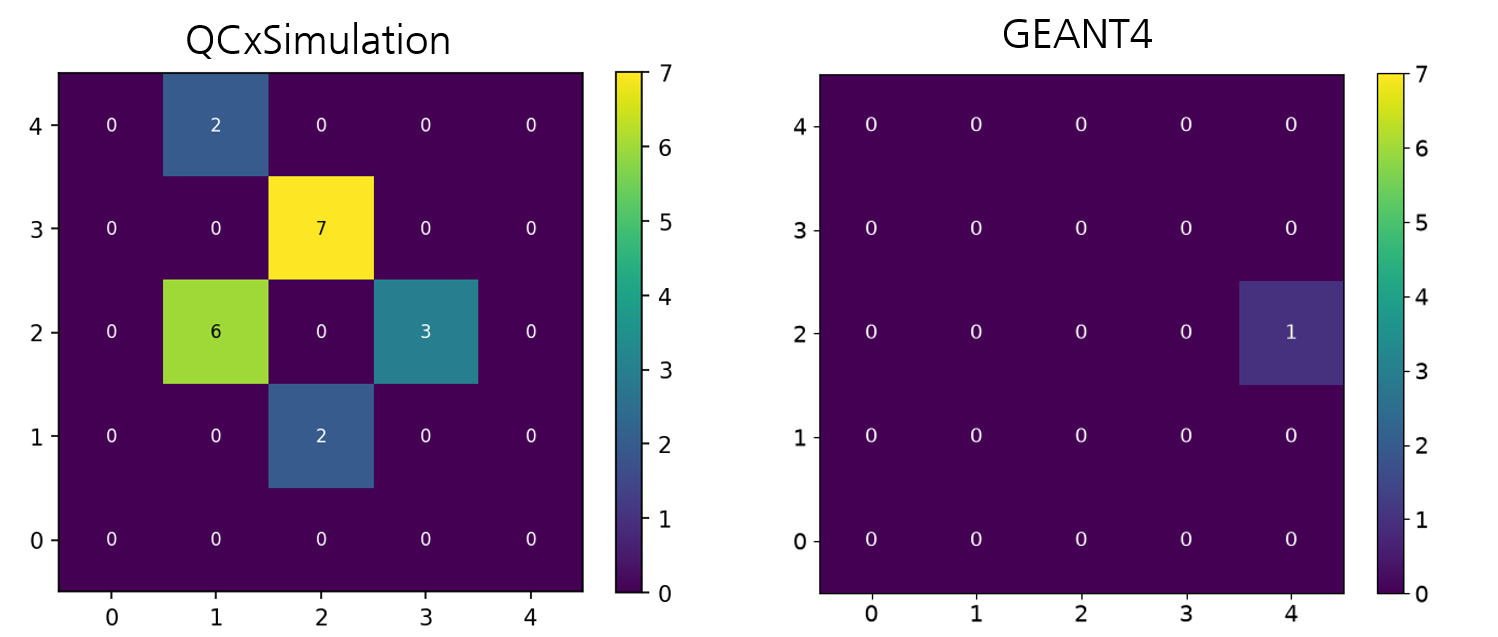}
        \caption{Use Case 2:~$80~\mathrm{keV}$}
        \label{fig:usecase2_geant1}
    \end{subfigure}
    \hfill
    \begin{subfigure}[b]{0.49\textwidth}
        \centering
        \includegraphics[width=\textwidth]{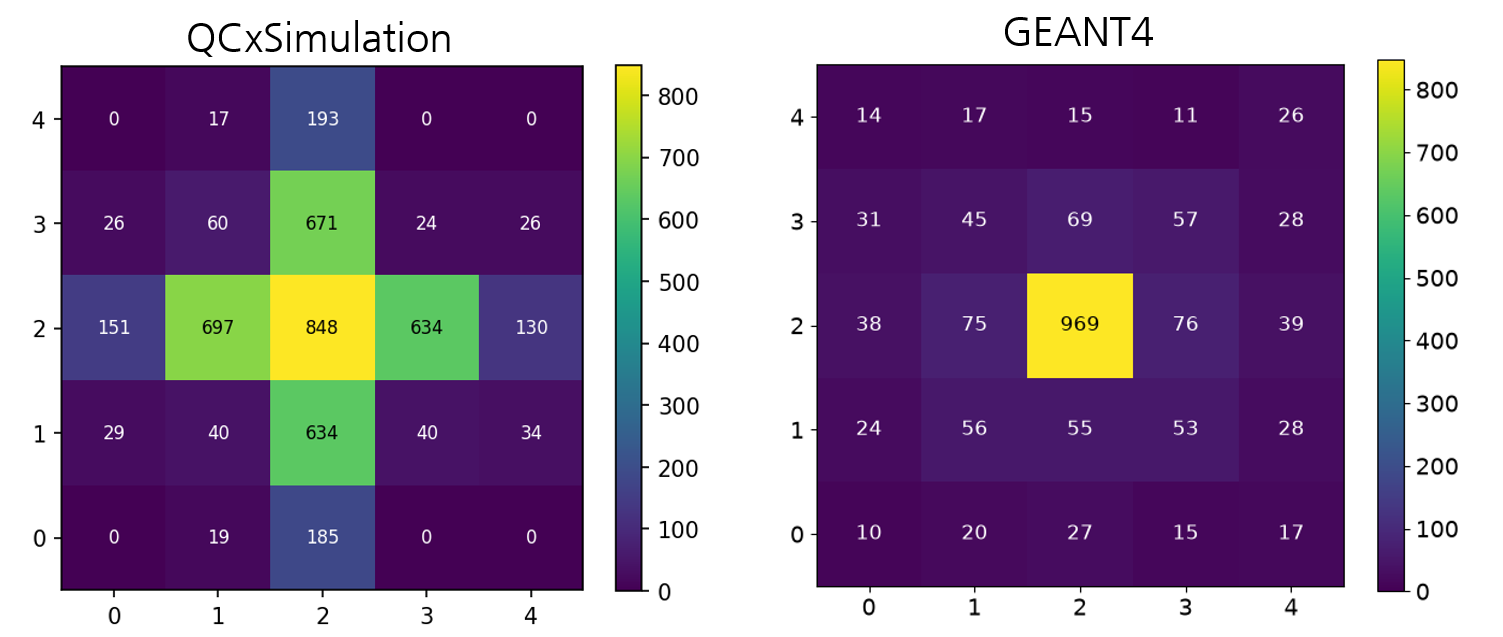}
        \caption{Use Case 2:~$220~\mathrm{keV}$}
        \label{fig:usecase2_geant2}
    \end{subfigure}

    % Second row: one subfigure centered
    \vspace{0.15cm}
    \begin{subfigure}[b]{0.49\textwidth}
        \centering
        \includegraphics[width=\textwidth]{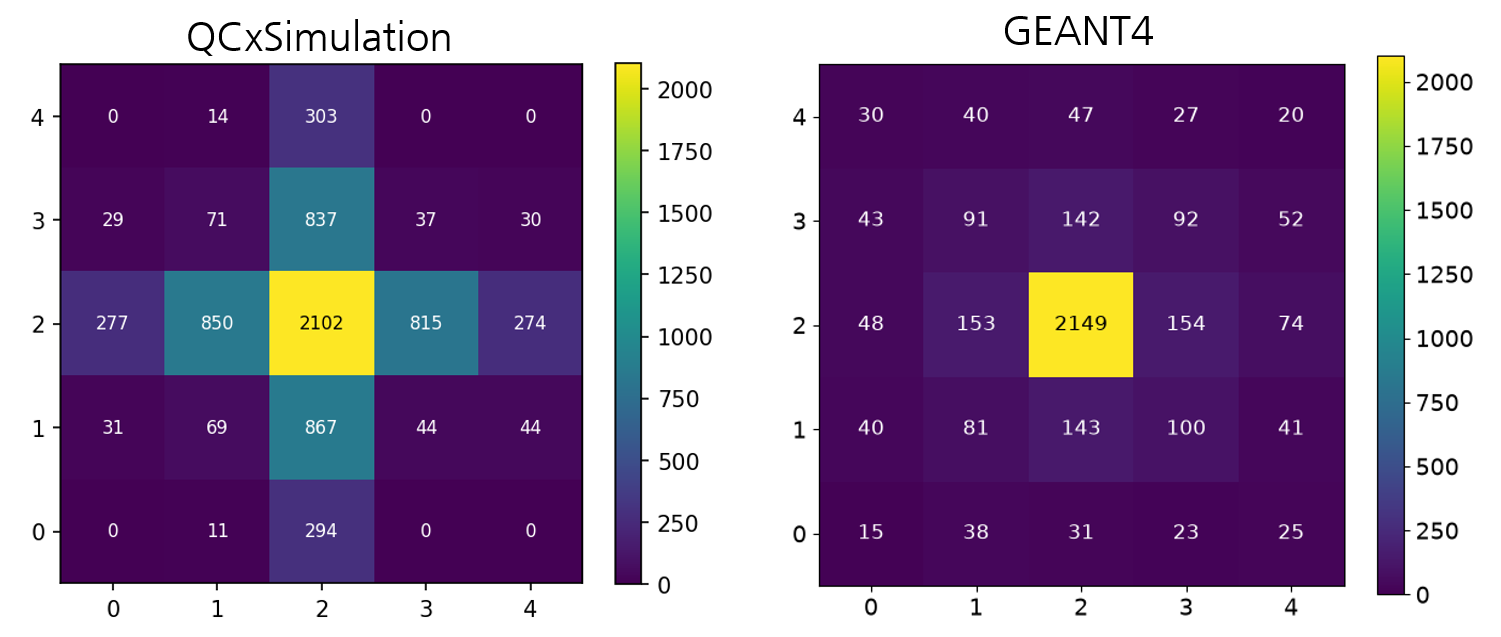}
        \caption{Use Case 2:~$400~\mathrm{keV}$}
        \label{fig:usecase2_geant3}
    \end{subfigure}

    \caption{Photon count of detector projections for Use Case 2: Comparison of QCxSimulation results \textit{all-detected} with the results from GEANT4}
    \label{fig:usecase2_geant4}
\end{figure}

% GEANT Comparison
\autoref{fig:usecase2_geant4} compares the \textit{all-detected} detector projections of the quantum walk simulation and the GEANT4 reference across all three photon energies. The two approaches are very similar. Any remaining differences are due to the physical discretization required to model the physics of continuous transport in a finite quantum circuit. \autoref{fig:usecase2_geant1} shows this comparison at $80~\mathrm{keV}$. At this energy, photoelectric absorption dominates in copper, so nearly all primary photons are absorbed within the $2~\mathrm{cm}$ copper block and only a small number of scattered photons reach the detector. \autoref{fig:usecase2_geant2} shows the comparison at $220~\mathrm{keV}$. Here, Compton scattering becomes dominant, allowing a greater fraction of primary photons to transmit through the copper block and producing a broader lateral scatter distribution. \autoref{fig:usecase2_geant3} shows the comparison at $400~\mathrm{keV}$. Compton scattering now dominates for both materials, yielding the highest primary transmission and a more uniform scatter distribution across the detector.

\subsection{Use Case 3}
\begin{figure}[htbp]
    \centering
    % First row: one subfigure centered
    \begin{subfigure}[b]{\textwidth}
        \centering
        \includegraphics[width=0.49\textwidth]{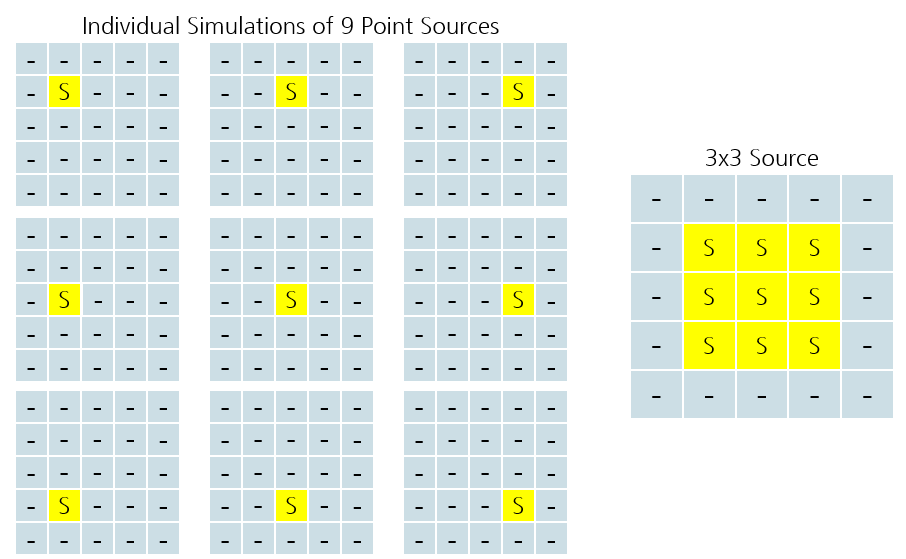}
        \caption{Individual point sources that have been combined, compared to the $3 \times 3$ source}
        \label{fig:usecase3_intro}
    \end{subfigure}
    
    % Second row: two subfigures side by side
    \vspace{0.15cm}
    \begin{subfigure}[b]{0.49\textwidth}
        \centering
        \includegraphics[width=\textwidth]{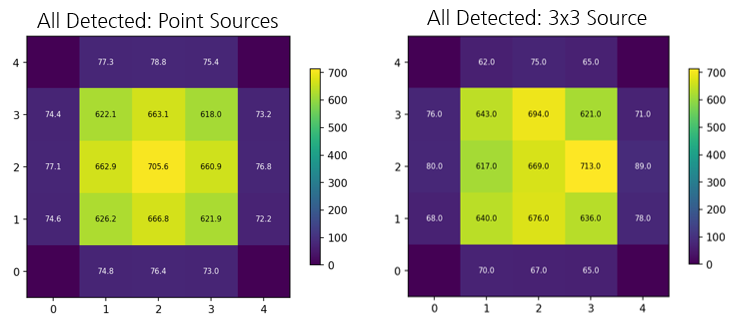}
        \caption{Photon count of detector projections for the summed individual projections and the $3\times 3$ result}
        \label{fig:usecase3_results}
    \end{subfigure}
    \hfill
    \begin{subfigure}[b]{0.49\textwidth}
        \centering
        \includegraphics[width=\textwidth]{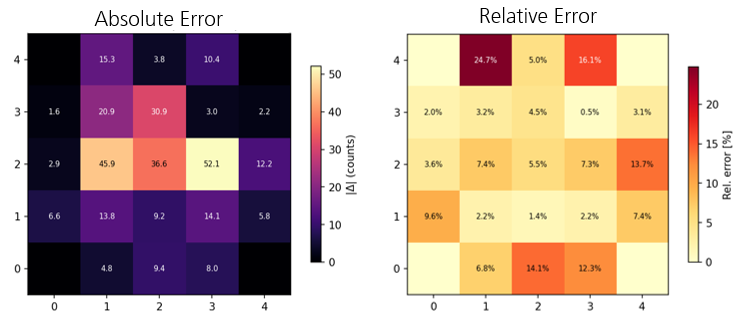}
        \caption{Absolute and relative errors of compared photon counts}
        \label{fig:usecase3_error}
    \end{subfigure}
    \caption{Superposition study for Use Case 3: comparison of the normalised sum of nine individual point-source simulations with a single 3×3 source run.}
    \label{fig:usecase3}
\end{figure}

% see the simulations in QCxSimulation\superposition_study
% superposition
The third use case tests whether the quantum walk correctly implements the principle of linear superposition, as in classical simulation~\cite{agostinelli_geant4simulation_2003}. The experiment consists of ten simulations, all using the aluminium phantom of Use Case~\ref{sec:usecase1} at $80~\mathrm{keV}$. Nine of these runs each employ a single point source placed at one of the nine positions $(x,y,4)$ for $x,y \in \{1,2,3\}$. Together, these position form a $3\times 3$ source. The tenth run uses the $3 \times 3$ source, which initialises the photon state as a uniform superposition over all nine positions simultaneously. After execution, the detector projections from the nine individual runs are summed and normalised by a factor of nine, in order to be compared with the result of the $3 \times 3$ run. \autoref{fig:usecase3_intro} shows the nine different point sources and the $3 \times 3$ source.
In classical Monte Carlo simulations, each photon path is independent. Therefore, a multi-pixel source image is the sum of single-pixel images, requiring a separate ray trace for each position. In contrast, a quantum circuit can encode all source positions in superposition. If implemented correctly, it exhibits the same additive behaviour. Therefore, the normalised sum of nine point-source runs should match the $3 \times 3$ source result. Any significant difference indicates unwanted quantum interference or operator errors.
%
% describing the result images
\autoref{fig:usecase3} shows the detector projections for both the normalised sum of nine individual point-source simulations and the single $3 \times 3$ source run. Both images show the same spatial structure: the central $3\times3$ detector region records the highest photon counts, while the surrounding air pixels remain dark. The absolute and relative errors (cf. \autoref{fig:usecase3_error}) show no systematic spatial pattern. The pixel-wise differences are consistent with statistical shot noise. Each individual point-source run uses $N=10,000$ shots at a single position, the $3 \times 3$ run distributes the same $N$ shots uniformly across all nine source positions, giving each position only $N/9 \approx 1,111$ effective shots. The dominant noise therefore comes from the $3 \times 3$ run, with an expected relative uncertainty of $3/\sqrt{N} \approx 3\%$~\cite{Knoll2010}. 
The high relative errors observed at the surrounding air pixels are statistical artefacts, as these pixels receive only a small number of scattered photons. With only a few photons reaching these pixels, even a difference of just one photon can result in a high relative error of several percent. Averaged over all pixels that receive primary photons, the relative error remains below the expected shot-noise uncertainty, confirming that no systematic discrepancy exists between the two approaches.

\section{Discussion}
The three use cases presented in Section~\ref{sec:results} demonstrate that the quantum walk algorithm successfully reproduces the dominant spatial and energetic features of X-ray radiographic transport. The results confirm that the proposed circuit is flexible enough to handle qualitatively different source geometries, multi-material phantoms, and spectral ranges. The following discussion addresses the physical accuracy of the model, its scalability and adaptability, and the pathway towards execution on real quantum hardware.

\subsection{Physical Accuracy and Modelling Approximations}
% To what extent does the algorithm deviate from reality?
    % Overshoot through z=-2
    % light strength of photons is divded to sum to 1
The quantitative comparison with GEANT4 simulations shows that quantum walks correctly captures the dominant radiographic behaviour. The observed deviations are the expected consequences of physical discretisation, which is necessary to map continuous transport physics onto a finite quantum circuit. The direction register encodes 14 discrete propagation directions, providing a coarse but suitable sampling of the forward-peaked Klein-Nishina angular distribution at various energies. 
The four-level energy discretisation captures the dominant energy changes that result from large-angle Compton scattering. However, it treats forward-like scattering, which causes minimal energy loss, as energy-neutral. Although this simplification introduces a systematic underestimation of energy loss, this remains small and predictable. The energy loss per forward-like scattering event is physically limited by the Compton formula and cannot increase indefinitely. The discretisation error per deflected scattering, which is the rounding error introduced by assigning a continuous energy value to one of four discrete levels, follows the quantisation bound $\delta E_{max}= \frac{1}{2}  \lvert E_{k+1} - E_{k}\rvert$~\cite{Knoll2010}. So, the continuous energy can deviate from its nearest discrete level by at most half of the distance to the next level. Both effects are limited in total because a photon can only undergo a finite number of scatters before being absorbed or leaving the grid.
Although the approximations result in slight differences in the intensities of the scattering patterns compared to GEANT4’s continuous models, the overall spatial structure of the projections remains unchanged. This demonstrates that the chosen degree of discretisation is effective in capturing the most significant physical processes.

A further design consideration is how the quantum state is normalised. The quantum state must always satisfy $\sum_k |\alpha_k|^2 = 1$, meaning that each measurement yields exactly one photon outcome. Consequently, the total number of detectable photons is restricted by the number of measurement shots. For example, $10,000$ shots will result in a maximum of $10,000$ detected photons. When using multiple source positions encoded via superposition, the shot budget is shared between these positions. Each position receives approximately $N_{shots}$ / $N_{positions}$ effective measurements. For instance, a $3 \times 3$ source distributes the $10,000$ shots among 9 source positions, giving each position only $\sim 1,111$ shots. Therefore, activating more source positions does not by itself increase the total detector counts. Instead, the total probability is distributed among all active positions, meaning that each position contributes fewer photons proportionally. This is not a fundamental difference between the quantum and classical approaches, since contributions from different source positions add linearly in both cases. The difference arises from the normalization. In classical simulation, more source positions increase the total photon count only if the number of photons per position is kept fixed; otherwise, a fixed total photon budget is simply redistributed across the source positions, as in the quantum case.

An additional design feature specific to the direction discretisation is the forward-leap optimisation introduced by the directions $d \in \{10,11,12,13\}$ with $\Delta z=-2$. These directions represent small-angle forward scattering ($\theta \approx 26.6^\circ$) and enable photons to traverse the grid more efficiently. This reduces the minimum number of steps required to reach the detector, enabling additional possible paths in the configurations discussed. For instance, it allows photons to hit the four neighbouring pixels around the central pixel $(2,2)$.
The trade-off is that photons at layer $z=1$ with $\Delta z=-2$ overshoot the detector and are therefore killed by the boundary logic. This geometric constraint does not affect any of the phantom configurations studied.

\subsection{Scalability and Quantum Advantage}
% Algorithm can be adapted to different grid sizes, different materials, different sources, more directions, more energy levels -- current limitation is the number of qubits
The algorithm can be adapted to diverse simulation scenarios. The grid can be scaled by extending the position register. Meanwhile, the material composition can be updated by reassigning the voxel index sets $V_m$ and re-computing the xraylib look-up tables. The source positions can be set arbitrarily over one or more source planes, enabling a range of geometries from pencil to cone beams. Direction and energy resolution can be improved by adding qubits to the respective registers, yielding finer angular and spectral resolution.
The circuit complexity scales as $\mathcal{O}(S \cdot N_{dir}^{2} \cdot \log(N_{x}N_{y}N_{z}))$. This scaling arises from $S$ walk steps, each of which applies a full coin-and-shift cycle. The coin gate encodes the scattering probability amplitude of the $N_{dir}$ incoming directions to each of the $N_{dir}$ outgoing directions, resulting in $N_{dir}^2$. The shift gate requires a position register with a size of $\log(N_{x}N_{y}N_{z})$ qubits. The current implementation requires $18 + N_{mat}$ qubits with a total gate count of $\mathcal{O}(10^3)$.

All simulations in this work were executed on a classical statevector simulator~\cite{javadi-abhari_quantum_2024}.
To verify the numerical reproducibility of the simulation, ten independent runs of scenario \autoref{fig:usecase1_intro}~(a) under identical conditions were compared by computing the coefficient of variation (CV)~\cite{brown_coefficient_1998} of key photon counts, including total detected, primary, and scattered photons. The mean CV of approximately $2\%$ across all runs is consistent with the expected statistical shot noise of $\frac{1}{N} \approx 1\%$ at $N = 10,000$ shots, confirming that the observed variability is attributable due to stochastic sampling~\cite{Knoll2010}.

The current circuit depth and qubit count exceed the capabilities of current NISQ devices.
Nevertheless, the quantum walk formulation has two concrete advantages which will become feasible as fault-tolerant hardware advances. The first of these is a structural advantage in computational scaling. In a classical, higher-order scattering simulation, every possible scattering history must be calculated individually. At each scattering event, the photon can proceed in up to $N_{dir}$ directions. This causes the total number of paths to grow exponentially with scatter order $S$, as $\mathcal{O}(N_{dir}^{S})$. The quantum walk avoids this entirely by encoding all admissible photon paths as a superposition in a single quantum state. This means that a single application of the coin-and-shift operator advances every possible scattering trajectory in parallel, while keeping the total circuit cost polynomial. 
The second advantage relates to statistical precision. The current workflow uses a standard shot-based measurement. Here, the number of photons arriving at each detector pixel is estimated by running the circuit $N$ times and counting the results. The accuracy of this estimate improves by a factor of only $\mathcal{O}(\frac{1}{\sqrt{N}})$, so increasing the number of measurements by a factor of four only doubles the precision. Quantum amplitude estimation (QAE)~\cite{montanaro_quantum_2016} could instead extract the same result with $\mathcal{O}(\frac{1}{N})$ convergence, requiring quadratically fewer measurement repetitions for a given target accuracy. QAE amplifies the amplitude of the target measurement through quantum interference prior to measurement. This concentrates the statistical information into a single phase that can be extracted with exponential precision via phase estimation.
However, QAE requires the entire quantum walk to execute as a single, unitary operation, meaning no measurements may occur during the walk itself. This is incompatible with the current implementation, where a classical bit is written at every walk step to record whether a photon scattered, which partially collapses the quantum state mid-execution. Integrating QAE would require keeping all scattering information inside the quantum circuit, i.e. in quantum superposition, instead of extracting it step by step via measurements. This would substantially increase the depth of the circuit beyond what is currently feasible on the state vector simulator, where the memory requirements already increase exponentially with the number of qubits.
The advantages of polynomial scaling and QAE only become feasible on fault-tolerant devices. The gate error rates of current NISQ devices cause noise to accumulate and overwhelm the physical signal at the required circuit depths. Crucially, these advantages are most significant promising where classical simulation struggles the most: with large phantoms, high scatter orders and fine angular discretisations. This makes the quantum walk a compelling future alternative in scenarios where classical methods are least practical.

\section{Conclusion}
This work presented a quantum walk-based X-ray simulation. The circuit successfully reproduces the energy-dependent angular distributions of first- and second-order Compton scattering via Klein-Nishina weighted look-up tables, as well as multi-material contrast that has been validated against GEANT4 simulations. All simulations were executed on a classical statevector simulator, as current NISQ hardware remains insufficient. This work establishes a physically validated quantum walk framework for X-ray simulation. As fault-tolerant hardware becomes more readily available, this framework will provide a compelling alternative to classical Monte Carlo methods.

\bibliographystyle{vancouver}
\bibliography{references}

@incollection{brown_coefficient_1998,
	address = {Berlin, Heidelberg},
	title = {Coefficient of {Variation}},
	isbn = {9783642803307 9783642803284},
	url = {http://link.springer.com/10.1007/978-3-642-80328-4_13},
	language = {en},
	urldate = {2026-07-06},
	booktitle = {Applied {Multivariate} {Statistics} in {Geohydrology} and {Related} {Sciences}},
	publisher = {Springer Berlin Heidelberg},
	author = {Brown, Charles E.},
	collaborator = {Brown, Charles E.},
	year = {1998},
	doi = {10.1007/978-3-642-80328-4_13},
	pages = {155--157},
}

@article{montanaro_quantum_2016,
	title = {Quantum algorithms: an overview},
	volume = {2},
	issn = {2056-6387},
	shorttitle = {Quantum algorithms},
	url = {https://www.nature.com/articles/npjqi201523},
	doi = {10.1038/npjqi.2015.23},
	language = {en},
	number = {1},
	urldate = {2026-07-06},
	journal = {npj Quantum Information},
	author = {Montanaro, Ashley},
	month = jan,
	year = {2016},
	pages = {15023},
}

@misc{javadi-abhari_quantum_2024,
	title = {Quantum computing with {Qiskit}},
	url = {http://arxiv.org/abs/2405.08810},
	doi = {10.48550/arXiv.2405.08810},
	urldate = {2026-07-06},
	publisher = {arXiv},
	author = {Javadi-Abhari, Ali and Treinish, Matthew and Krsulich, Kevin and Wood, Christopher J. and Lishman, Jake and Gacon, Julien and Martiel, Simon and Nation, Paul D. and Bishop, Lev S. and Cross, Andrew W. and Johnson, Blake R. and Gambetta, Jay M.},
	month = jun,
	year = {2024},
	note = {arXiv:2405.08810},
}

@article{Klein1929,
	title = {Über die {Streuung} von {Strahlung} durch freie {Elektronen} nach der neuen relativistischen {Quantendynamik} von {Dirac}},
	volume = {52},
	number = {11},
	journal = {Zeitschrift für Physik},
	author = {Klein, Oskar and Nishina, Yoshio},
	year = {1929},
	pages = {853--868},
}

@book{Knoll2010,
	address = {Hoboken, N.J},
	edition = {4th ed},
	title = {Radiation detection and measurement},
	isbn = {9780470131480},
	publisher = {John Wiley},
	author = {Knoll, Glenn F.},
	year = {2010},
}

@article{Brunetti2004,
	title = {A library for {X}-ray–matter interaction cross sections for {X}-ray fluorescence applications},
	volume = {59},
	copyright = {https://www.elsevier.com/tdm/userlicense/1.0/},
	issn = {05848547},
	url = {https://linkinghub.elsevier.com/retrieve/pii/S0584854704001843},
	doi = {10.1016/j.sab.2004.03.014},
	language = {en},
	number = {10-11},
	urldate = {2026-06-29},
	journal = {Spectrochimica Acta Part B: Atomic Spectroscopy},
	author = {Brunetti, A. and Sanchez Del Rio, M. and Golosio, B. and Simionovici, A. and Somogyi, A.},
	month = oct,
	year = {2004},
	pages = {1725--1731},
}

@article{born_quantenmechanik_1926,
	title = {Quantenmechanik der {Stoßvorgänge}},
	volume = {38},
	copyright = {http://www.springer.com/tdm},
	issn = {1434-6001, 1434-601X},
	url = {http://link.springer.com/10.1007/BF01397184},
	doi = {10.1007/BF01397184},
	language = {de},
	number = {11-12},
	urldate = {2026-06-29},
	journal = {Zeitschrift für Physik},
	author = {Born, Max},
	month = nov,
	year = {1926},
	pages = {803--827},
}

@inproceedings{Mosier2023,
	address = {Sydney NSW Australia},
	title = {Quantum {Ray} {Marching}: {Reformulating} {Light} {Transport} for {Quantum} {Computers}},
	isbn = {9798400703157},
	shorttitle = {Quantum {Ray} {Marching}},
	url = {https://dl.acm.org/doi/10.1145/3610548.3618151},
	doi = {10.1145/3610548.3618151},
	language = {en},
	urldate = {2026-06-10},
	booktitle = {{SIGGRAPH} {Asia} 2023 {Conference} {Papers}},
	publisher = {ACM},
	author = {Mosier, Logan and Mcguire, Morgan and Hachisuka, Toshiya},
	month = dec,
	year = {2023},
	pages = {1--9},
}

@article{Santos2025,
	title = {Towards {Quantum} {Ray} {Tracing}},
	volume = {31},
	copyright = {https://ieeexplore.ieee.org/Xplorehelp/downloads/license-information/IEEE.html},
	issn = {1077-2626, 1941-0506, 2160-9306},
	url = {https://ieeexplore.ieee.org/document/10494551/},
	doi = {10.1109/TVCG.2024.3386103},
	number = {4},
	urldate = {2026-06-10},
	journal = {IEEE Transactions on Visualization and Computer Graphics},
	author = {Santos, Luís Paulo and Bashford-Rogers, Thomas and Barbosa, João and Navrátil, Paul},
	month = apr,
	year = {2025},
	pages = {2223--2234},
}

@inproceedings{lanzagorta_hybrid_2005,
	address = {Los Angeles, California},
	title = {Hybrid quantum-classical computing with applications to computer graphics},
	copyright = {https://www.acm.org/publications/policies/copyright\_policy\#Background},
	url = {http://portal.acm.org/citation.cfm?doid=1198555.1198723},
	doi = {10.1145/1198555.1198723},
	language = {en},
	urldate = {2026-06-10},
	booktitle = {{ACM} {SIGGRAPH} 2005 {Courses} on   - {SIGGRAPH} '05},
	publisher = {ACM Press},
	author = {Lanzagorta, Marco and Uhlmann, Jeffrey K.},
	year = {2005},
	pages = {2},
}

@inproceedings{Alves2019,
	address = {Faro, Portugal},
	title = {A {Quantum} {Algorithm} for {Ray} {Casting} using an {Orthographic} {Camera}},
	copyright = {https://ieeexplore.ieee.org/Xplorehelp/downloads/license-information/IEEE.html},
	isbn = {9781728163789},
	url = {https://ieeexplore.ieee.org/document/8955061/},
	doi = {10.1109/ICGI47575.2019.8955061},
	urldate = {2026-06-10},
	booktitle = {2019 {International} {Conference} on {Graphics} and {Interaction} ({ICGI})},
	publisher = {IEEE},
	author = {Alves, Carolina and Santos, Luis Paulo and Bashford-Rogers, Thomas},
	month = nov,
	year = {2019},
	pages = {56--63},
}

@misc{Devkota2025,
	title = {{QuaRT}: {A} toolkit for the exploration of quantum methods for radiation transport},
	shorttitle = {{QuaRT}},
	url = {http://arxiv.org/abs/2511.12356},
	doi = {10.48550/arXiv.2511.12356},
	urldate = {2026-06-10},
	publisher = {arXiv},
	author = {Devkota, Rasmit and Wise, John H.},
	month = nov,
	year = {2025},
	note = {arXiv:2511.12356},
}

@article{Lu2023,
	title = {Improved quantum supersampling for quantum ray tracing},
	volume = {22},
	issn = {1573-1332},
	url = {https://link.springer.com/10.1007/s11128-023-04114-x},
	doi = {10.1007/s11128-023-04114-x},
	language = {en},
	number = {10},
	urldate = {2026-06-10},
	journal = {Quantum Information Processing},
	author = {Lu, Xi and Lin, Hongwei},
	month = sep,
	year = {2023},
	pages = {359},
}

@misc{Lu2022,
	title = {A {Framework} for {Quantum} {Ray} {Tracing}},
	url = {http://arxiv.org/abs/2203.15451},
	doi = {10.48550/arXiv.2203.15451},
	urldate = {2026-06-10},
	publisher = {arXiv},
	author = {Lu, Xi and Lin, Hongwei},
	month = mar,
	year = {2022},
	note = {arXiv:2203.15451},
}

@inproceedings{Zhang2025,
	address = {Albuquerque, NM, USA},
	title = {Towards {Hybrid} {Spatial} {Structure} {Rendering} {Models} {Using} {Quantum} {Orthographic} {Ray} {Casting}},
	copyright = {https://doi.org/10.15223/policy-029},
	isbn = {9798331557362},
	url = {https://ieeexplore.ieee.org/document/11250325/},
	doi = {10.1109/QCE65121.2025.00261},
	urldate = {2026-06-10},
	booktitle = {2025 {IEEE} {International} {Conference} on {Quantum} {Computing} and {Engineering} ({QCE})},
	publisher = {IEEE},
	author = {Zhang, Zirui},
	month = aug,
	year = {2025},
	pages = {2410--2418},
}

@misc{Aharonov2003,
	title = {A {Simple} {Proof} that {Toffoli} and {Hadamard} are {Quantum} {Universal}},
	url = {http://arxiv.org/abs/quant-ph/0301040},
	doi = {10.48550/arXiv.quant-ph/0301040},
	urldate = {2026-06-10},
	publisher = {arXiv},
	author = {Aharonov, Dorit},
	month = jan,
	year = {2003},
	note = {arXiv:quant-ph/0301040},
}

@book{von_neumann_mathematische_1932,
	title = {Mathematische {Grundlagen} der {Quantenmechanik}},
	url = {https://gdz.sub.uni-goettingen.de/id/PPN379400774?tify=%7B%22view%22%3A%22info%22%7D},
	publisher = {Springer},
	author = {von Neumann, Johann},
	year = {1932},
}

@article{reiter_simct_2016,
	title = {{SimCT}: a simulation tool for {X}-ray imaging},
	volume = {21},
	url = {https://www.ndt.net/?id=18746},
	number = {2},
	urldate = {2026-05-10},
	journal = {e-Journal of Nondestructive Testing},
	author = {Reiter, Michael and Erler, Marco and Kuhn, Christoph and Gusenbauer, Christian and Kastner, Johann},
	year = {2016},
}

@article{bellon_artist_2007,
	title = {{aRTist} - analytical {Radiographic} {Testing} inspection simulation tool},
	volume = {10},
	url = {https://www.ndt.net/?id=4917},
	urldate = {2026-05-10},
	journal = {e-Journal of Nondestructive Testing},
	author = {Bellon, Carsten and Jaenisch, Gerd-Rüdiger},
	year = {2007},
}

@article{tabary_sindbad_2007,
	title = {Sindbad - a realistic multi-purpose and scalable {X}-ray simulation tool for {NDT} applications},
	volume = {10},
	url = {https://www.ndt.net/?id=4920},
	urldate = {2026-05-10},
	journal = {e-Journal of Nondestructive Testing},
	author = {Tabary, Joachim and Hugonnard, Patrick and Mathy, Francoise},
	year = {2007},
}

@article{neffati_novi-sim_2023,
	title = {Novi-{Sim}: {A} fast {X}-ray tomography simulation software for laboratory and synchrotron systems to generate training databases for deep learning applications},
	volume = {28},
	issn = {14354934},
	shorttitle = {Novi-{Sim}},
	url = {https://www.ndt.net/search/docs.php3?id=27739},
	doi = {10.58286/27739},
	number = {3},
	urldate = {2026-05-10},
	journal = {e-Journal of Nondestructive Testing},
	author = {Neffati, Dajla and Autret, Awen and Berheas, Tom and Jayde, Livingstone and Fayard, Barbara},
	month = mar,
	year = {2023},
}

@article{oconnell_hybrid_2025,
	title = {Hybrid {GPU} {Monte} {Carlo} software toolkit for fast and accurate cone‐beam {CT} simulation},
	volume = {52},
	issn = {0094-2405, 2473-4209},
	url = {https://aapm.onlinelibrary.wiley.com/doi/10.1002/mp.18011},
	doi = {10.1002/mp.18011},
	language = {en},
	number = {8},
	urldate = {2026-05-10},
	journal = {Medical Physics},
	author = {O'Connell, Jericho and Jacobson, Matthew and Harris, Thomas and Bruegger, Raphael and Ferguson, Dianne and Fueglistaller, Rony and Hu, Yue‐Houng and Morf, Daniel and Birrer, Vera and Arroyo, Pablo Corral and Lehmann, Mathias and Myronakis, Marios and Berbeco, Ross},
	month = aug,
	year = {2025},
	pages = {e18011},
}

@phdthesis{schielein_analytische_2018,
	type = {{PhD} thesis},
	title = {Analytische {Simulation} und {Aufnahmeplanung} für die industrielle {Röntgencomputertomographie}},
	doi = {https://nbn-resolving.org/urn:nbn:de:bvb:20-opus-169236},
	school = {Universität Würzburg},
	author = {Schielein, Richard},
	year = {2018},
}

@article{gallio_gpu_2015,
	title = {A {GPU} {Simulation} {Tool} for {Training} and {Optimisation} in {2D} {Digital} {X}-{Ray} {Imaging}},
	volume = {10},
	issn = {1932-6203},
	url = {https://dx.plos.org/10.1371/journal.pone.0141497},
	doi = {10.1371/journal.pone.0141497},
	language = {en},
	number = {11},
	urldate = {2026-05-10},
	journal = {PLOS ONE},
	author = {Gallio, Elena and Rampado, Osvaldo and Gianaria, Elena and Bianchi, Silvio Diego and Ropolo, Roberto},
	editor = {Chen, Chin-Tu},
	month = nov,
	year = {2015},
	pages = {e0141497},
}

@misc{vidal_simulation_2009,
	title = {Simulation of {X}-ray {Attenuation} on the {GPU}},
	url = {http://diglib.eg.org/handle/10.2312/LocalChapterEvents.TPCG.TPCG09.025-032},
	doi = {10.2312/LOCALCHAPTEREVENTS/TPCG/TPCG09/025-032},
	language = {en},
	urldate = {2026-05-10},
	publisher = {The Eurographics Association},
	author = {Vidal, Franck P. and Garnier, Manuel and Freud, Nicolas and Létang, Jean Michel and John, Nigel W.},
	year = {2009},
}

@article{badal_accelerating_2009,
	title = {Accelerating {Monte} {Carlo} simulations of photon transport in a voxelized geometry using a massively parallel graphics processing unit},
	volume = {36},
	copyright = {http://onlinelibrary.wiley.com/termsAndConditions\#vor},
	issn = {0094-2405, 2473-4209},
	url = {https://aapm.onlinelibrary.wiley.com/doi/10.1118/1.3231824},
	doi = {10.1118/1.3231824},
	language = {en},
	number = {11},
	urldate = {2026-05-10},
	journal = {Medical Physics},
	author = {Badal, Andreu and Badano, Aldo},
	month = nov,
	year = {2009},
	pages = {4878--4880},
}

@article{anagnostou_effect_2026,
	title = {The {Effect} of {Scatter} {Radiation} on {Image} {Resolution} in {Gridless} {Portable} {X}-{Ray} {Imaging}: {A} {Monte} {Carlo} {Study}},
	volume = {16},
	issn = {2076-3417},
	shorttitle = {The {Effect} of {Scatter} {Radiation} on {Image} {Resolution} in {Gridless} {Portable} {X}-{Ray} {Imaging}},
	url = {https://www.mdpi.com/2076-3417/16/7/3152},
	doi = {10.3390/app16073152},
	language = {en},
	number = {7},
	urldate = {2026-05-09},
	journal = {Applied Sciences},
	author = {Anagnostou, Ilias and Liaparinos, Panagiotis and Michail, Christos and Valais, Ioannis and Fountos, George and Kandarakis, Ioannis and Kalyvas, Nektarios},
	month = mar,
	year = {2026},
	pages = {3152},
}

@article{andriiashen_quantifying_2023,
	title = {Quantifying the effect of {X}-ray scattering for data generation in real-time defect detection},
    journal={Journal of X-ray Science and Technology},
	copyright = {arXiv.org perpetual, non-exclusive license},
	url = {https://arxiv.org/abs/2305.12822},
	doi = {10.48550/ARXIV.2305.12822},
	urldate = {2026-05-09},
	author = {Andriiashen, Vladyslav and van Liere, Robert and van Leeuwen, Tristan and Batenburg, K. Joost},
	year = {2023},
}

@article{hassan_hybrid_2026,
	title = {Hybrid multimodal artificial intelligence, vision sensory, and robotic cyber-physical systems for plastic inspection and sorting in digital circular remanufacturing: {A} review},
	volume = {3},
	copyright = {https://creativecommons.org/licenses/},
	issn = {3041-0746, 3029-2573},
	shorttitle = {Hybrid multimodal artificial intelligence, vision sensory, and robotic cyber-physical systems for plastic inspection and sorting in digital circular remanufacturing},
	url = {https://accscience.com/journal/IJAMD/3/1/10.36922/IJAMD025340030},
	doi = {10.36922/IJAMD025340030},
	number = {1},
	urldate = {2026-05-09},
	journal = {International Journal of AI for Materials and Design},
	author = {Hassan, Syed Ali and Beliatis, Michail J.},
	month = mar,
	year = {2026},
	pages = {1},
}

@article{metiner_deep_2025,
	title = {Deep {Learning}-{Enhanced} {X}-{Ray} {Computed} {Tomography} for {Defect} {Detection} in {Composite} {Structures}},
	volume = {44},
	issn = {0195-9298, 1573-4862},
	url = {https://link.springer.com/10.1007/s10921-025-01268-9},
	doi = {10.1007/s10921-025-01268-9},
	language = {en},
	number = {4},
	urldate = {2026-05-09},
	journal = {Journal of Nondestructive Evaluation},
	author = {Metiner, Abdullah and Nikishkov, Yuri and Makeev, Andrew and Koçyiğit, Mustafa T.},
	month = dec,
	year = {2025},
	pages = {127},
}

@article{velleman_state---art_2026,
	title = {State-of-the-{Art} {Medical} {Imaging} {Supporting} {Supermicrosurgery}: {A} {Scoping} {Review}, {Practical} {Guidelines}, and {Illustrative} {Case} {Report}},
	volume = {53},
	copyright = {https://creativecommons.org/licenses/by/4.0/},
	issn = {2234-6163, 2234-6171},
	shorttitle = {State-of-the-{Art} {Medical} {Imaging} {Supporting} {Supermicrosurgery}},
	url = {http://www.thieme-connect.de/DOI/DOI?10.1055/a-2798-0198},
	doi = {10.1055/a-2798-0198},
	language = {en},
	number = {02},
	urldate = {2026-05-09},
	journal = {Archives of Plastic Surgery},
	author = {Velleman, Jos and Kwon, Jin Geun and Pak, Changsik John and Suh, Hyunsuk Peter and Hong, Joon Pio},
	month = mar,
	year = {2026},
	pages = {178--190},
}

@article{luo_artificial_2026,
	title = {Artificial intelligence and multimodal imaging in orthopaedics: from technological advances to clinical translation},
	volume = {12},
	issn = {2296-858X},
	shorttitle = {Artificial intelligence and multimodal imaging in orthopaedics},
	url = {https://www.frontiersin.org/articles/10.3389/fmed.2025.1728248/full},
	doi = {10.3389/fmed.2025.1728248},
	urldate = {2026-05-09},
	journal = {Frontiers in Medicine},
	author = {Luo, Guangan and Tan, Shuanglong and Luo, Lincong and Hu, Konghe},
	month = jan,
	year = {2026},
	pages = {1728248},
}

@article{compton_quantum_1923,
	title = {A {Quantum} {Theory} of the {Scattering} of {X}-rays by {Light} {Elements}},
	volume = {21},
	copyright = {http://link.aps.org/licenses/aps-default-license},
	issn = {0031-899X},
	url = {https://link.aps.org/doi/10.1103/PhysRev.21.483},
	doi = {10.1103/PhysRev.21.483},
	language = {en},
	number = {5},
	urldate = {2026-05-09},
	journal = {Physical Review},
	author = {Compton, Arthur H.},
	month = may,
	year = {1923},
	pages = {483--502},
}

@article{strutt_xv_1871,
	title = {{XV}. {On} the light from the sky, its polarization and colour},
	volume = {41},
	issn = {1941-5982, 1941-5990},
	url = {https://www.tandfonline.com/doi/full/10.1080/14786447108640452},
	doi = {10.1080/14786447108640452},
	language = {en},
	number = {271},
	urldate = {2026-05-09},
	journal = {The London, Edinburgh, and Dublin Philosophical Magazine and Journal of Science},
	author = {Strutt, J.W.},
	month = feb,
	year = {1871},
	pages = {107--120},
}

@article{makarov_computed_2023,
	title = {Computed tomography with or without radiation},
	volume = {28},
	issn = {14354934},
	url = {https://www.ndt.net/search/docs.php3?id=27745},
	doi = {10.58286/27745},
	number = {3},
	urldate = {2026-05-09},
	journal = {e-Journal of Nondestructive Testing},
	author = {Makarov, Jurij and Basting, Melanie and Schielein, Richard and Schmitt, Michael and Kabitzky, Dominik and Sukowski, Frank and Voland-Salamon, Virginia},
	month = mar,
	year = {2023},
}

@article{auer_mesh_2023,
	title = {Mesh modeling of system geometry and anatomy phantoms for realistic {GATE} simulations and their inclusion in {SPECT} reconstruction},
	volume = {68},
	issn = {0031-9155, 1361-6560},
	url = {https://iopscience.iop.org/article/10.1088/1361-6560/acbde2},
	doi = {10.1088/1361-6560/acbde2},
	number = {7},
	urldate = {2026-05-09},
	journal = {Physics in Medicine \& Biology},
	author = {Auer, Benjamin and Könik, Arda and Fromme, Timothy J and De Beenhouwer, Jan and Kalluri, Kesava S and Lindsay, Clifford and Furenlid, Lars R and Kuo, Philip H and King, Michael A},
	month = apr,
	year = {2023},
	pages = {075015},
}

@article{agostinelli_geant4simulation_2003,
	title = {Geant4—a simulation toolkit},
	volume = {506},
	issn = {01689002},
	url = {https://linkinghub.elsevier.com/retrieve/pii/S0168900203013688},
	doi = {10.1016/S0168-9002(03)01368-8},
	language = {en},
	number = {3},
	urldate = {2026-05-07},
	journal = {Nuclear Instruments and Methods in Physics Research Section A: Accelerators, Spectrometers, Detectors and Associated Equipment},
	author = {Agostinelli, S. and Allison, J. and Amako, K. and Apostolakis, J. and Araujo, H. and Arce, P. and Asai, M. and Axen, D. and Banerjee, S. and Barrand, G. and Behner, F. and Bellagamba, L. and Boudreau, J. and Broglia, L. and Brunengo, A. and Burkhardt, H. and Chauvie, S. and Chuma, J. and Chytracek, R. and Cooperman, G. and Cosmo, G. and Degtyarenko, P. and Dell'Acqua, A. and Depaola, G. and Dietrich, D. and Enami, R. and Feliciello, A. and Ferguson, C. and Fesefeldt, H. and Folger, G. and Foppiano, F. and Forti, A. and Garelli, S. and Giani, S. and Giannitrapani, R. and Gibin, D. and Gómez Cadenas, J.J. and González, I. and Gracia Abril, G. and Greeniaus, G. and Greiner, W. and Grichine, V. and Grossheim, A. and Guatelli, S. and Gumplinger, P. and Hamatsu, R. and Hashimoto, K. and Hasui, H. and Heikkinen, A. and Howard, A. and Ivanchenko, V. and Johnson, A. and Jones, F.W. and Kallenbach, J. and Kanaya, N. and Kawabata, M. and Kawabata, Y. and Kawaguti, M. and Kelner, S. and Kent, P. and Kimura, A. and Kodama, T. and Kokoulin, R. and Kossov, M. and Kurashige, H. and Lamanna, E. and Lampén, T. and Lara, V. and Lefebure, V. and Lei, F. and Liendl, M. and Lockman, W. and Longo, F. and Magni, S. and Maire, M. and Medernach, E. and Minamimoto, K. and Mora De Freitas, P. and Morita, Y. and Murakami, K. and Nagamatu, M. and Nartallo, R. and Nieminen, P. and Nishimura, T. and Ohtsubo, K. and Okamura, M. and O'Neale, S. and Oohata, Y. and Paech, K. and Perl, J. and Pfeiffer, A. and Pia, M.G. and Ranjard, F. and Rybin, A. and Sadilov, S. and Di Salvo, E. and Santin, G. and Sasaki, T. and Savvas, N. and Sawada, Y. and Scherer, S. and Sei, S. and Sirotenko, V. and Smith, D. and Starkov, N. and Stoecker, H. and Sulkimo, J. and Takahata, M. and Tanaka, S. and Tcherniaev, E. and Safai Tehrani, E. and Tropeano, M. and Truscott, P. and Uno, H. and Urban, L. and Urban, P. and Verderi, M. and Walkden, A. and Wander, W. and Weber, H. and Wellisch, J.P. and Wenaus, T. and Williams, D.C. and Wright, D. and Yamada, T. and Yoshida, H. and Zschiesche, D.},
	month = jul,
	year = {2003},
	pages = {250--303},
}

\end{document}